\documentclass[12pt,aps,pre,showpacs]{revtex4}

\usepackage{latexsym}
\usepackage{amssymb}
\usepackage{epsfig}
\usepackage{subfigure}
\usepackage{times}

\def\div{\mathop{\rm div}\nolimits}
\def\D{D}

\def\H{H}
\def\K{G}
\def\P{p}
\def\G{\Gamma}

\def\Ylm{\mathrm{Y}_{lm}}

\begin{document}

\title{Hexatic undulations in curved geometries}
\author{Peter Lenz}
\altaffiliation[Current address: ]{Institut Curie, UMR 168, 26 rue
d'Ulm, F-75248 Paris C\'edex 05, France}
\affiliation{Lyman Laboratory of Physics, Harvard
  University, Cambridge, MA 02138}
\author{David R. Nelson}
\affiliation{Lyman Laboratory of Physics, Harvard
  University, Cambridge, MA 02138}
\date{\today }

\begin{abstract}
We discuss the influence of two-dimensional hexatic order on capillary
waves and undulation modes in spherical and cylindrical geometries.
In planar geometries, extended bond-orientational order has only a
minor effect on the fluctuations of liquid surfaces or lipid bilayers.
However, in curved geometries, the long-wavelength spectrum of these
ripples is altered. We calculate this frequency shift and discuss
applications to spherical vesicles, liquid metal droplets, bubbles and
cylindrical jets coated with surface-active molecules, and to
multielectron bubbles in liquid helium at low temperatures.  Hexatic
order also leads to a shift in the threshold for the fission
instability of charged droplets and bubbles, and for the
Plateau-Rayleigh instability of liquid jets.
\end{abstract}

\maketitle

\section{Introduction}

In two dimensions the melting from a crystal to an isotropic liquid
can be a two-stage process \cite{halp78}, driven by the sequential
unbinding of dislocations \cite{halp78,kost73} and disclinations
\cite{halp78}.  At low temperatures $T<T_m$ dislocations are
suppressed due to their cost in elastic energy. However, their free
energy decreases with increasing temperature.  At a temperature
$T=T_m$, the quasi-long-ranged translational order of the crystal is
destroyed by the dissociation of dislocation pairs. This transition
leads to an intervening hexatic phase, which still exhibits extended
orientational correlations.  The unbinding of disclination pairs sets
in at a higher temperature $T=T_i$.  In this second transition the
quasi-long-ranged orientational order of the hexatic phase is
destroyed, leading to an isotropic liquid.  This mechanism allows the
melting transition to be continuous in contrast to the first-order
melting (directly to an isotropic fluid) predicted by Landau
\cite{land37}.

Several experimental systems have illuminated the nature of
two-dimensional melting. A nearly ideal system is electrons on helium
\cite{grim79,glat88,devi84}. The electrons are trapped on the surface
of liquid helium by a submerged, positively charged capacitor plate.
Their separations are rarely less than 1000\AA, so the in-plane
physics is that of classical particles with a repulsive $1/r$
potential. The liquid helium does not freeze at low temperatures, so
it is possible to cool the electrons on this liquid substrate well
below a sharply defined 2d freezing temperature ($T_m \sim 0.5K$)
\cite{grim79}. On the theoretical side, important parameters such as
the 2d shear modulus and dislocation core energy are easily calculated
with this potential \cite{fish79}. Computer simulations \cite{morf79}
reveal a shear modulus which appears to drop to zero at the melting
temperature. Measurements of the shear modulus \cite{devi84} and
specific heat \cite{glat88} are {\em consistent} with a continuous
dislocation mediated melting transition. However, in these experiments
it is difficult to determine experimentally if hexatic order and a
disclination unbinding transition are in fact present above $T_m$.

However, experimental evidence for hexatic order {\em has} been found
in a variety of other systems, including free standing liquid crystal
films \cite{chou98} and Langmuir-Blodgett surfactant monolayers
\cite{knob92}.  A hexatic-to-liquid transition has been observed in
two-dimensional magnetic bubble arrays \cite{sesh91}. Furthermore,
there is strong support for two-stage continuous melting from recent
experiments on two-dimensional colloidal crystals
\cite{murr92,zahn99}. In this case, the colloids can be directly
imaged by video-microscopy, thus allowing a precise test of the theory.

The modest time scales available even on the fastest computers make
equilibration in Monte Carlo or molecular dynamics simulations of
two-dimensional melting difficult. However, there is now evidence via
computer simulations for continuous melting and a narrow sliver of
hexatic phase for hard disks \cite{jast99} and for particles
interacting with a repulsive $1/r^{12}$ potential \cite{bagc96}. There
are also indications of defect mediated melting transitions for the
familiar Lennard-Jones 6-12 pair potential \cite{some98}.

In the above experiments, order was typically probed via diffraction
or by direct measurement of correlation functions in real space. It is
difficult to use these methods, however, when hexatic order is present
on a curved surface, such as a sphere or a cylinder.  Examples where
hexatic and crystalline order might be present on a sphere include
"liposomes", i.e., closed vesicles composed of lipid bilayers
\cite{park92,evan96}, the surface of liquid metal droplets confined in
Paul traps \cite{davi97}, and multielectron bubbles submerged in
liquid helium \cite{leid95}.

Hexatic order in spherical liposomes seems likely because flat
two-dimensional planar layers of lipids such as DMPC 
(Dimyristoyl phosphatidylcholine) 
\cite{chia95}, similar to free standing liquid crystal films
\cite{huan92}, can exist in a variety of states with different degrees
of positional and orientational order. Examples include fluid, smectic
C, hexatic and crystalline phases.

Curved hexatic order may also arise on droplets.  Celestini et al.
\cite{cele97} have found evidence from computer simulations for
extended orientational correlations at the flat surface of supercooled
heavy noble liquid metals, such as Au, Pt, or Ir.  These metals have a
general tendency to reduce the interatomic distance at the surface.
Upon supercooling this effect is enhanced, leading, generally to a
two-dimensional crystalline surface layer. Under suitable cooling
conditions, hexatic order can also appear on the free surface of
undercooled liquid metals \cite{cele97}. Surface hexatic order may
also occur on water droplets coated with surface-active molecules,
given that there are already extensive observations of this type of
order in Langmuir-Blodgett monolayers \cite{knob92}.

Finally, one might expect spherical hexatic and crystalline order in
multielectron bubbles in helium. These arise when the helium surface
undergoes an electrohydrodynamic instability at high electron
densities. The surface then develops a regular array of dimples, each
containing $10^6$ electrons or more. As the electric field increases
these dimples deepen until eventually electrons break through the
interface.  After subduction, large numbers of electrons ($10^5-10^7$)
then coat the inside wall of a large (10-100$\mu$m radius) sphere of
helium vapor.  These multielectron bubbles have been observed to move
through the helium after their creation above a protruding anode
\cite{albr92,leid95,leid97}.  They are stable at low electron
densities since the Coulomb pressure can be compensated by the surface
tension of the helium liquid-vapor interface.  However, if the
electron density becomes too high the electrostatic repulsion exceeds
the balancing force and the bubble undergoes fission
\cite{salo81,land}.

One might hope that bond-orientational order in a flat membrane or
interface could be detected by its effects on the dynamics of
undulation modes or capillary waves. Unfortunately, hexatic order
couples only to the {\em Gaussian} curvature \cite{nels87}, which
vanishes for a simple sine wave deformation of a flat membrane or an
interface (cf. Eqs.~(\ref{eq:n11}) and (\ref{eq:2.38}) below). The
situation is different, however, when these excitations are
superimposed on a nontrivial background geometry such as that of a
sphere or a cylinder. In a recent short communication \cite{lenz01} we
have determined the effect of hexatic order on the undulation modes
and capillary wave excitations for the spherical systems described
above.  The frequency shift is large for liposomes with hexatic order.
Observable effects could also occur for liquid metal droplets,
surfactant coated water drops and in multielectron bubbles in helium.
In this paper we describe hexatic dynamics on spherical surfaces in
detail and extend the theory to include cylindrical geometries.

{\em Cylindrical} geometries could be realized by, e.g., coating a
liquid jet with a hexatic monolayer. These jets will (similar to
conventional liquids) undergo the well-known Plateau-Rayleigh shape
instability as soon as their length reaches a critical size.
However, the stiffness associated with extended orientational
correlations shifts the threshold of this instability and alters the
decay of the cylinder into a chain of droplets.  Cylinders provide
also an example where hexatic order is perfectly compatible with the
underlying geometry: since the Gaussian curvature of the cylinder
vanishes, no disclination defects are present in the ground state to
complicate the analysis.

The remainder of this paper is organized as follows. First, we analyze
the influence of surface hexatic order on spheres by considering
liquid droplets (Sect.~\ref{sec:3}). Then, in Sect.~\ref{sec:6} we
apply our analysis to cylindrical geometries.  Here, we concentrate on
liquid jets and we determine the effect of hexatic order on the
Plateau-Rayleigh instability. Next, we compute the shifted instability
due to hexatic order in multielectron bubbles (Sect.~\ref{sec:4}).  As
a last application we investigate in Sect.~\ref{sec:2} the influence
of hexatic order on the undulations of spherical vesicles.  Finally,
in Sect.~\ref{sec:7} we elaborate in detail on experimental
consequences of our work.

\section{Dynamics of liquid droplets with hexatic order}
\label{sec:3}

\begin{figure}
  \begin{center}
    \mbox{\epsfxsize=10cm    \epsffile{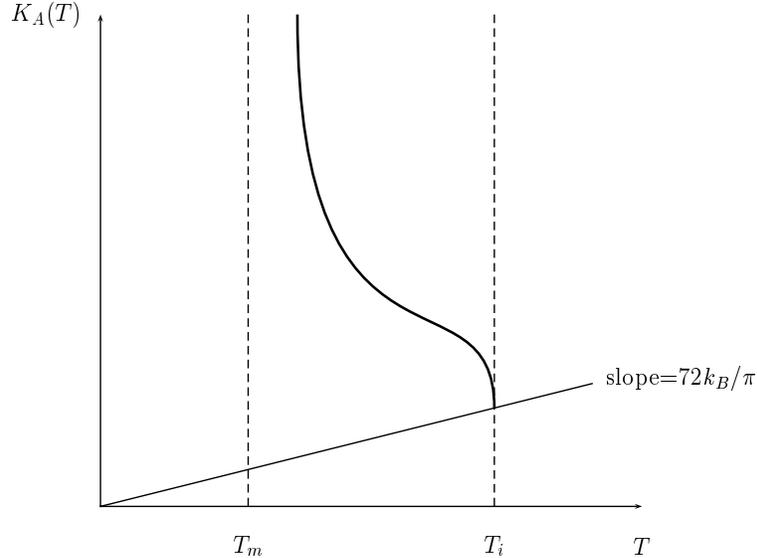}}
  \end{center}
\caption{\label{fig:1}
The hexatic stiffness $K_A$ as function of the temperature $T$. $K_A$
diverges near $T_m$ and jumps discontinuously to zero at $T_i$ with
$\lim_{T \rightarrow T_i^-}K_A(T)=72k_BT/\pi$  \cite{halp78,nels83b}.}
\end{figure}

We first discuss liquid droplets with surface hexatic order. The
equilibrium shape minimizes a droplet free energy $F_d$ given by
contributions from an interfacial energy and the hexatic degrees of
freedom 
\begin{equation}
    F_d =F_i+F_h \equiv 
\sigma \int dA
 + \frac{1}{2} K_A \int dA  \ D _in^j D ^i n_j,
\label{eq:3.1}
\end{equation}
where we use the summation convention throughout and $\sigma$ denotes
the surface tension of the interface of the liquid droplet.  For a
general manifold with internal coordinates $x=(x^1,x^2)$, the surface
element is given by $dA=\sqrt{g(x)} d^2x$, where $g(x)$ is the
determinant of the metric tensor $g_{ij}(x)$. For an undeformed sphere
with radius $R_0$, $x \equiv (\theta,\varphi)$ with polar coordinates
$\theta$ and $\varphi$ and $dA=R_0^2 \sin\theta d\theta d \varphi$.
The quantity $\vec{n}$ is a unit vector in the tangent plane with $n_i
n^i=1$ which identifies (modulo $2 \pi/6$) the long-range correlations
in the hexatic bond directions \cite{nels87}.  Here, $D_i n^j \equiv
g^{jk}D_in_k$, where $g^{ij}$ is the inverse of $g_{ij}$. The operator
$D_i$ denotes a covariant derivative with respect to the metric
$g_{ij}$. Thus, $D_in^j \equiv \partial _i n^j+\Gamma_{ki}^jn^k$,
where the $\Gamma_{ki}^j$ are Christoffel symbols of second kind. See, 
e.g., \cite{hild}. Close to the melting temperature $T_m$, the hexatic
stiffness $K_A \sim E_c (\xi_T/a_0)^2$, 
where $\xi_T$ is the translational correlation length, $a_0$ is the
particle spacing and $E_c$ is the dislocation core energy
\cite{halp78}. The ratio $K_A/k_BT$ jumps from an universal value
$72/\pi$ to zero when the hexatic melts into an isotropic liquid at
$T=T_i^-$ (see Fig.~\ref{fig:1}) \cite{halp78}.

Droplets have a nearly constant volume $V$ and the corresponding
constraint could be included in the free energy $F_d$. However, here it
is easier to first consider shape fluctuations which explicitly keep
the volume fixed.  Finally, in Eq. (\ref{eq:3.1}) we neglect effects
arising from gravity since we only consider droplets with radii $R_0
\ll l_c$ much smaller than the capillary length $l_c$, which is of
order millimeters or more for typical droplets in the earth's
gravitational field.

In the following we will consider the dynamics of shape fluctuations
of spherical droplets. As the droplet deforms its geometrical
properties change, i.e.,  the metric and the mean and Gaussian curvature
are altered.  The free energy (\ref{eq:3.1}) thus has a functional
dependence on the underlying droplet shape. This dependence can be
treated most efficiently by parameterizing the surface of the droplet
by its surface vector $\vec{R}$. For a sphere, we have
\begin{equation}
  \label{eq:2.2}
\vec{R}(x^1,x^2)\equiv\vec{R}(\theta,\varphi) =R_0(\theta,\varphi)
\left( \sin \theta 
  \cos \varphi, \sin \theta 
  \sin \varphi, \cos \theta\right).
\end{equation}
See Appendix~\ref{app:1} for important geometrical quantities such as
the metric tensor $g_{ij}(x)$, the mean curvature $H(x)$ and the
Gaussian curvature $\K(x)$ in terms of the surface vector
$\vec{R}(x^1,x^2)$. 

The fundamental assumption which underlies the hexatic free energy
discussed above is that the configuration of minimal elastic energy
corresponds to a vector field $\vec{n}_0$ where $\vec{n}_0(x+dx)$ can
be obtained from $\vec{n}_0(x)$ by parallel transport of $\vec{n}_0$.
On a sphere however, curvature introduces ``frustration'' since
parallel transport of $\vec{n}$ along closed loops on the surface
leads to a rotation of $\vec{n}$.  Because of this frustration the
ground state of hexatic order on a sphere has at least 12 positive
disclinations. This constraint is a consequence of the Poincar\'e
index theorem \cite{fran}, which states that a vector field on a
surface with genus $g$ and Euler characteristic $E=2(1-g)$ must have
singularities with total vorticity $2 \pi E$.  As a consequence, order
which is identified by a vector order-parameter field on a curved
geometry frustrated by a nonzero integrated Gaussian curvature
\begin{equation}
\overline{\K}  = \int dA G(x),
\end{equation}
always has topological defects \cite{nels83}. On a sphere with $g=0$
and $E=2$ hexatics must have a minimum of 12 defects with charges $2
\pi/6$ \cite{lube87}.

The energy of an isolated disclination in a hexatic diverges
logarithmically in flat space. However, this energy is reduced due to
screening by the Gaussian curvature of the sphere.  This point can be
made more precise by introducing a local bond-angle field $\theta$,
the angle between $\vec{n}$ and some local reference frame. The
transverse part of $\theta$ is then connected with the disclination
density \cite{nels87,nels83}. As shown in Appendix~\ref{app:2}, the
elastic free energy associated with hexatic order can then be written
as \cite{bowi00}
\begin{equation}
\label{eq:4}
    F_h  = -\frac{1}{2} K_A \int dA \int dA'
 \left [\K(x)-s(x) \right] 
\G(x,x')
 \left [\K(x')-s(x') \right] . \label{eq:2.4}
\end{equation}
Here, $\G(x,x')$ is the inverse Laplacian on the sphere (see Appendix~
\ref{app:2}), and $s(x)$ the disclination density \cite{nels83},
\begin{equation}
\label{eq:2.5}
s(x) \equiv 
\frac{1}{ \sqrt{g(x)}} \sum_{i=1}^{N_d} q_i \delta(x-x_i),
\end{equation}
with $N_d$ disclinations of charge $q_i=\pm 2 \pi/6$ at positions
$x_i$.  In deriving Eq. (\ref{eq:2.4}) we have assumed that the
regular part of the bond-angle field $\theta^{reg}$ relaxes rapidly on
the time scale of shape deformations. In Sect.~\ref{sec:2} we will
show that this assumption is indeed justified for the systems
considered here.  Finally, it should be emphasized that Eqs.
(\ref{eq:2.4}) and (\ref{eq:2.5}) hold for arbitrary geometries, not
just for that of the sphere (cf. Appendix~\ref{app:2}).

The defects minimize $F_h$ by arranging themselves to approximately
match the Gaussian curvature, which is $\K(x)\equiv1/R_0^2$ for a
rigid sphere of radius $R_0$. Deep in a hexatic phase on a sphere, we
expect $N_d=12$, corresponding to 12 fivefold disclinations which lie
on the vertices of an icosahedron. With polar coordinates such that
there are disclinations at the north and south pole, the 12 defect
locations entering Eq. (\ref{eq:2.5}) are given by
\begin{equation}
\label{eq:2.6}
  (\theta_k,\varphi_k) \in \left \{(0,0),(\gamma,\frac{2 \pi k}{5})_{0
      \leq 
      k \leq 4}, (\pi-\gamma,\frac{\pi}{5}+\frac{2 \pi k}{5})_{0 \leq
      k \leq 4},(\pi,0)  \right \},
\end{equation}
where
\begin{equation}
 \gamma \equiv \cos^{-1} \frac{1}{\sqrt{5}}. 
\end{equation}

\subsection{Fluctuation spectrum}

\label{sec:2.1}

\begin{figure}
  \begin{center}
    \mbox{\epsfxsize=8cm    \epsffile{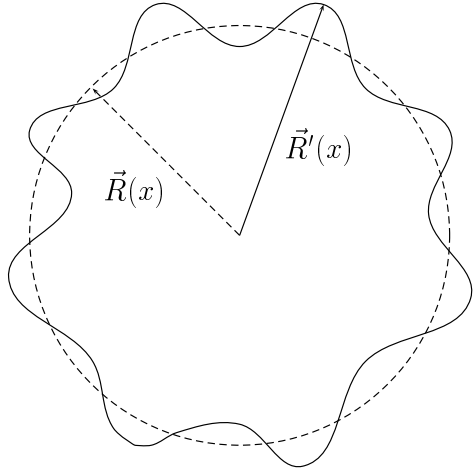}}
  \end{center}
\caption{\label{fig:2} Parameterization of the droplet shape by the
  surface vector $\vec{R}$. The deformed surface can be parameterized
  by $\vec{R}'(x,t)=R_0 \left [1+\zeta(x,t) \right ] \vec{N}$, where
  $\vec{N}$ is the unit normal.}
\end{figure}

To investigate the influence of hexatic order on spherical droplets, we
study deformations about the equilibrium configuration. We expand the
free energy $F_d$ in a small time-dependent displacement field 
$\delta\vec{R}(x,t)$, where 
\begin{equation}
\label{eq:2.8}
  \vec{R}'(x,t)=\vec{R}_0(x)+\delta \vec{R}(x,t)
\end{equation}
is the deformed surface and $\vec{R}_0(x)$ is given by Eq.
(\ref{eq:2.2}). For liquid droplets with hexatic order it is
sufficient to consider purely normal displacement fields, cf. Fig.
\ref{fig:2}. Thus,
\begin{equation}
 \delta \vec{R}(x,t)=R_0 \zeta(x,t) \vec{N}(x) ,
\end{equation}
where $\vec{N}(x)=\vec{R}_0/R_0$ is the normal vector of the sphere.
The dimensionless function $\zeta$ can be expanded in terms of
spherical harmonics
\begin{equation}
\label{eq:10}
\zeta(x,t)=\sum_{l=0}^{\infty } \sum_{m=-l}^{l} r_{lm}(t)\Ylm(x).
\end{equation}

In the absence of defects, the expansion of $F_d$ in $\zeta$ would be
straightforward. On the sphere, however, one has to deal with 12 
discrete disclination charges which produce a small static
icosahedral surface deformation. We initially neglect this
discreteness and approximate these defects by a smeared 
distribution of defect ``charge''. As will be shown in
Sect.~\ref{sec:2.2},  corrections arising from the 
discrete nature of $s(x)$ are irrelevant for the oscillation
frequencies $\omega(l)$ with $l<6$.  These considerations can be made
more precise by expanding $s(x)$ in terms of spherical harmonics
\begin{equation}
\label{eq:2.11}
  s(x)=\K_0+\frac{1}{R_0^2} \sum_{l=1}^{\infty }\sum_{m=-l}^{l}
  s_{lm} \Ylm(x) ,  
\end{equation}
where $\K_0$ is the undeformed Gaussian curvature, $\K_0=1/R_0^2$, and
$s_{lm} \equiv \sum_{i=1}^{12}q_i{\mathrm{Y}_{lm}^*}(x_i)$. To obtain
the last equation, we have used the representation of the
$\delta$-function in terms of spherical harmonics,
\begin{equation}
  \delta(\cos\theta-\cos\theta') \delta(\varphi-\varphi') =
\sum_{l=0}^{\infty } \sum_{m=-l}^{l} \Ylm^* (\theta',\varphi')
  \Ylm(\theta ,\varphi).
\end{equation} 
In Eq. (\ref{eq:2.4}) we initially smear out the disclination charge by
setting 
$s_{lm}\simeq 0$ for all $l > 0$.

With this approximation, the hexatic order has no influence on the
droplet shape and the equilibrium configuration is a sphere with a
radius $R_0$.  However, the presence of hexatic order with a nonzero
stiffness constant $K_A$ nevertheless has an important effect on the
fluctuation spectrum $\omega(l)$ of a spherical droplet.

To calculate $\omega(l)$, we adopt the treatment of capillary waves on
spherical droplets without hexatic order \cite{land2} and consider 
the incompressible Navier-Stokes equation for the fluid of the droplet
\newcounter{tempnr} \setcounter{tempnr}{\value{equation}}
\stepcounter{tempnr} \setcounter{equation}{0}
\renewcommand{\theequation}{\arabic{tempnr}\alph{equation}}
\begin{eqnarray}
\label{eq:2.14}
\renewcommand{\theequation}{A\arabic{equation}a}
  \rho_l \frac{\partial \vec{v}}{\partial t} + \rho_l (\vec{v} \cdot
  \nabla ) 
  \vec{v}& = & -
 \nabla  p + \eta \nabla^2 \vec{v} \\
\renewcommand{\theequation}{\arabic{equation}b} 
\label{eq:2.14b}
\nabla  \cdot \vec{v} =0.
\end{eqnarray}
\setcounter{equation}{\value{tempnr}}
\renewcommand{\theequation}{\arabic{equation}}Here, $p$ is the
hydrostatic pressure in the presence of an interface and $\rho_l$,
$\eta$ and $\vec{v}$ denote the density, shear viscosity, and the
velocity of the liquid inside the droplet, respectively.

For droplets 
the inertial terms in the
Navier-Stokes equation (\ref{eq:2.14}) dominate and 
effects of viscosity are irrelevant. Upon neglecting the non-linear
term, the boundary conditions can be treated most efficiently
by introducing 
a  velocity potential $\Phi$ with 
$\vec{v}=\nabla \Phi$. Then, integration across the
droplet interface leads to 
\begin{equation}
\label{eq:3.2}
  \rho_l \left. \frac{\partial \Phi (r)}{\partial t} 
  \right |_{r=(R_0+\zeta R_0)^-} 
-  \rho_v \left. \frac{\partial \Phi (r)}{\partial t} 
  \right |_{r=(R_0+\zeta R_0)^+} 
= \Delta p (x) ,
\end{equation}
where $\rho_v$ is
the vapor density of the surrounding medium.  Eq.~(\ref{eq:3.2}) relates the
pressure difference between the inside and outside of the droplet to
the generalized pressure discontinuity $\Delta p(x)$ caused by the shape
displacement.  Since $\div \vec{v}=0$ one has $\nabla^2 \Phi=0$, a
Laplace equation with solutions of the form \cite{land2}
\begin{equation}
\label{eq:3.3}
  \Phi(r,x,t) = \left \{
    \begin{array}{lcl}
\sum_{l,m} A_{lm}^<(t) \Ylm(x) \left ( \frac{r}{R_0}\right
)^{l} &
\quad \mbox{ for } \quad & r<R_0(1+\zeta ) \\
\sum_{l,m} A_{lm}^>(t) \Ylm(x) \left ( \frac{R_0}{r}\right )^{l+1} &
\quad \mbox{ for } \quad & r>R_0(1+\zeta ) .
    \end{array}
 \right .
\end{equation}
The displacement field $\zeta$ and the velocity potential $\Phi$ are
related by the boundary condition that the interface velocity must
match the fluid velocity,
\begin{equation}
R_0  \dot{\zeta} \equiv R_0 \frac{ \partial\zeta}{\partial t}=
\left. \frac{ \partial \Phi}{\partial  r}\right |_{r=R_0+\zeta R_0}. 
\label{eq:3.4}
\end{equation}
Hence, the coefficients in Eq. (\ref{eq:3.3}) are given by 
\begin{equation}
\label{eq:3.17}
 A_{lm}^>(t)=-R_0^2 \frac{\dot{r}_{lm}(t)}{l+1}
\end{equation}
and
\begin{equation}
\label{eq:3.18}
  A_{lm}^<(t)=R_0^2 \frac{\dot{r}_{lm}(t)}{l}.
\end{equation}
Upon setting 
\begin{equation}
 \Delta p(x) \equiv \sum_{l,m} \Delta p_{lm}(t)\Ylm(x) \equiv -
 \sum_{l,m} 
 \frac{\delta F_d'(r_{lm})}{R_0^3 \delta r_{lm}^*} \Ylm(x),
\label{eq:2.23b}
\end{equation}
(where $r_{lm}^*$ denotes the complex conjugate of $ r_{lm}$ and
$F_d'$ the free energy of the deformed droplet), 
one then finds
\begin{equation}
\label{eq:3.52}
\Delta p_{lm}(t)=\left (
    \frac{\rho_l}{l}+\frac{\rho_v}{l+1}\right ) R_0^2
  \ddot{r}_{lm}(t). 
\end{equation}
Note the influence of hexatic order appears through the term 
$\delta F_d'/\delta r_{lm}^*$ in Eq. (\ref{eq:2.23b}).

To determine the coefficients $p_{lm}$ in Eq.
(\ref{eq:2.23b}), one has to calculate the variation of the droplet 
free energy.  Details about this calculation can be found 
in Appendix~\ref{app:4}. For small deviations from a spherical shape, 
the deformed droplet has volume $V'$ with
\begin{equation}
\label{eq:3.53p}
  V'-V=R_0^3 \left (\sqrt{4 \pi} r_{00}+\sum_{l=1}^{\infty }
    \sum_{m=-l}^{l} |r_{lm}|^2 \right ) + \mathcal{O}(r_{lm}^3),
\end{equation}
(see Eq. (\ref{eq:a4.5})). Thus, the volume constraint $V=V'$
appropriate to droplets can be
incorporated directly  by considering only displacements which fulfill 
(to leading order in the $r_{lm}$'s)
\begin{equation}
\label{eq:2.29}
  r_{00}= - \frac{1}{\sqrt{4 \pi}}\sum_{l=1}^{\infty } \sum_{m=-l}^{l}
  |r_{lm}|^2.  
\end{equation}
With the constraint of fixed volume, the interfacial contribution to
the free energy becomes 
\begin{equation}
\label{eq:3.53}
 F_{i}'=\sigma \int d^2x \sqrt{g+\delta g}=F_i+ \frac{1}{2}\sigma
 R_0^2\sum_{l=1}^{\infty } \sum_{m=-l}^{l}|r_{lm}|^2(l-1)(l+2) , 
\end{equation}
where $F_i=\sigma 4 \pi R_0^2$ is the interfacial free energy of the
undeformed droplet. Eq. (\ref{eq:3.53})
follows from Eqs. (\ref{eq:a4.4}) and 
(\ref{eq:2.29}). 
The hexatic free energy is given by
\begin{equation}
\label{eq:2.31}
 F_h'=\frac{1}{2}K_A   \sum_{l=1}^{\infty } \sum_{m=-l}^{l}
|r_{lm}|^2
  \frac{(l-1)^2(l+2)^2}{l(l+1)} ,
\end{equation}
cf. Eq. (\ref{eq:a4.23}). 

Upon setting $F_d'=F_i'+F_h'$ and 
\begin{equation}
\label{eq:3.25}
  r_{lm}=r^0_{lm}(t)e^{-i \omega(l)t} ,
\end{equation}
and evaluating Eq. (\ref{eq:2.23b}), we find for the fluctuation
spectrum (provided $l>0$ and $\rho_{v} \ll \rho_{l}$) \cite{footnew}
\begin{equation}
\label{eq:n10a}
  \omega^2=\frac{\sigma}{\rho_{l}R_0^3}l(l-1)(l+2)\left [
    1+\frac{K_A}{\sigma R_0^2} \frac{(l-1)(l+2)}{l(l+1)} \right ].
\end{equation}
Eq.~(\ref{eq:n10a})  shows that hexatic order only affects
undulation modes in a curved geometry: In the flat space limit of
large $R_0$ and $l \gg 1$ with $k \equiv l/R_0$ fixed, one has
\begin{equation}
\label{eq:n11}
  \omega^2 \simeq \frac{k^3}{\rho_l} \left [ \sigma+
    \frac{K_A}{R_0^2}\right ] ,
\end{equation}
which exhibits explicitely a hexatic correction to the usual capillary
wave spectrum.  The hexatic contribution, however, drops out as $R_0
\rightarrow \infty$ and we recover the result for capillary waves of a
flat fluid surface \cite{land2}. Thus, it is essential to study
deformations of a curved geometry to reveal the presence of hexatic
order.

In general, the undulation frequency Eq.~(\ref{eq:n10a}) depends on
the ratio $\frac{K_A}{\sigma R_0^2}$ which for hexatic order on
surfactant-coated water drops or at the surface of supercooled liquid
metal droplets is $K_A/\sigma R_0^2 \simeq (\xi_T/R_0)^2(E_c/\sigma
a_0^2)$.  This ratio grows to become of order 
$E_c/\sigma a_0^2 \simeq {\mathcal O}(1)$  close to a
continuous hexatic-to-crystal transition, i.e., when $\xi_T \approx
R_0$ (see also Sect.~\ref{sec:7}).

\subsection{Effect of defects on the spectrum}
\label{sec:2.2}

We now take the deformations associated with a discrete array of 12
disclination defects  into account,
i.e. we consider nonzero $s_{lm}$ with $l>0$ in Eq. (\ref{eq:2.11}). 
We first neglect the possibility of disclination motion. Thus, we
assume that on the time scale of the characteristic frequency 
(\ref{eq:n10a}), the disclinations remain in 
fixed   positions which minimize the hexatic free energy of the 
undeformed sphere. Thus, for $N_d=12$ disclinations at the vertices of an
icosahedron, the positions $x_i$ in
Eq. (\ref{eq:2.5}) are given by Eq. (\ref{eq:2.6}). 
As the sphere is deformed, the hexatic free energy then changes as (up 
to first order in $r_{lm}$)  
\begin{equation}
  \delta^{(1)}F_h=-K_A \sum_{l=1}^{\infty } \sum_{m=-l}^{l}
  \frac{(l-1)(l+2)}{l(l+1)} s_{lm} 
  r_{lm}^*. 
\end{equation}
This follows from Eq. (\ref{eq:a4.23}) since  $\delta s_{lm}$ is of
order $r_{lm}$ and the terms of order $s_{lm}\delta s_{lm}$ vanish
since the defect arrangement on the sphere minimizes $F_h$. 
To calculate the variation of the interfacial contribution the volume
constraint has to be included in the free energy. Thus, by 
considering the modified free energy 
\begin{equation}
  \tilde{F}_d=F_d+\int dV p(x), 
\end{equation}
(where $p \equiv p_{ex}-p_{in}$ is the pressure difference between
outside and inside of the droplet), the complete first variation
becomes
\begin{eqnarray}
  \delta^{(1)} \tilde{F}_d& = & 
 \sigma \int d^2x 
\sqrt{g(x)}   \delta^{(1)} g(x) + \int d^2x 
\sqrt{g(x)} p  \delta^{(1)} V(x)\nonumber \\ & & 
 - K_A \int d^2 x \sqrt{g(x)} \frac{1}{R_0^2}
\sum_{l=1}^{\infty } \sum_{m=-l}^{l} \frac{(l-1)(l+2)}{l(l+1)}
s_{lm}\Ylm(x) \sum_{l',m'}r^*_{l'm'} 
{\mathrm{Y}_{l'm'}^*}(x),
\end{eqnarray}
where $\delta^{(1)}g$ is given by Eq. (\ref{eq:a4.3}) and $\delta^{(1)}V$ by
Eq. (\ref{eq:a4.5}). 
The shape equation for quasi-spherical droplets with hexatic order
becomes 
\begin{equation}
\label{eq:2.36}
  p+2 \sigma H(x) =
 K_A  \frac{1}{R_0^3}
\sum_{l=1}^{\infty } \sum_{m=-l}^{l} \frac{(l-1)(l+2)}{l(l+1)}\Ylm
(x)s_{lm}. 
\end{equation}
Thus, nonzero coefficients $s_{lm}$ affect the mean curvature 
$H(x)$ of the stationary droplet. This equation simplifies upon making
the ansatz
\begin{equation}
\label{eq:2.37}
 H(x)=H_0+\widetilde{\delta H}(x) \equiv H_0+ \sum_{l=1}^{\infty }
\sum_{m=-l}^{l}h_{lm}\Ylm(x), 
\end{equation}
with
\begin{equation}
  2H_0 \sigma +p=0.
\end{equation}
One then finds the
extremal equation for the mean curvature, namely, 
\begin{equation}
  2H(x)=2H_0+\frac{K_A}{\sigma R^2_0} \frac{1}{R_0}
  \sum_{l=1}^{\infty } \sum_{m=-l}^{l}
  \frac{(l-1)(l+2)}{l(l+1)}s_{lm}\Ylm(x) .
\end{equation}
On the other hand, a deformed sphere has mean curvature 
(see Eq. (\ref{eq:a4.14}))
\begin{equation}
\label{eq:2.25}
2H'(x)=\frac{2}{R_0}+ \frac{1}{R_0} \sum_{l,m} (l-1)(l+2) r_{lm}\Ylm(x). 
\end{equation}

\begin{figure}
  \begin{center}
    \mbox{\epsfxsize=5cm  \epsfclipon     \epsffile{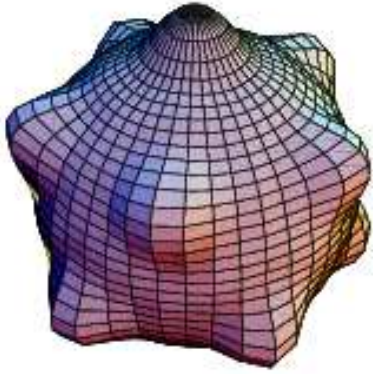}}
  \end{center}
\caption{\label{fig3}
Shape of a liquid droplet with hexatic order and 12 disclinations
lying on the vertices of an
icosahedron (surface deformations associated with the defects 
are exaggerated).}
\end{figure}

Comparison of the last two equations then leads immediately to the 
static surface deformation coefficients 
\begin{equation}
 r_{lm}^0=s_{lm}\frac{K_A}{\sigma R_0^2 l(l+1)}, 
\end{equation}
for $l>0$ and $r_{lm}^0=0$ for $l=m=0$
in the ground state.  Thus, for $K_A \neq 0$ the defects deform the
droplet as indicated in Fig.~\ref{fig3}.  However, nonzero $s_{lm}$
have no influence on the frequencies $\omega(l)$ for $0<l<6$. Indeed,
icosahedral symmetry insures that $s_{lm}=0$, and hence $r_{lm}^0=0$,
unless $l=6,10,12,...$ \cite{nels89}, so corrections of order $s_{lm}$ have no
influence on the frequencies $\omega(l)$ for small $l$. Thus, provided
the positions of the disclinations remain fixed on the time scale of
an undulation (i.e. $\delta s_{lm}=0$ holds in Eq.  (\ref{eq:a4.23}))
the dispersion relation (\ref{eq:n10a}) remains valid for $0<l<6$.
Because disclination motion is catalyzed by absorption and emission of
dislocations with mean spacing $\xi_T$, the disclination diffusion
constant is $D_5 \approx (a_0/\xi_T)^2D_0$, where the $D_0$ is the particle
diffusion constant. For a surfactant coated water droplet of radius
$R_0=1\mathrm{mm}$ one has (for $\sigma=10^{-3}\mathrm{N/m}$)
$\omega(l=2)\simeq 100 \mathrm{Hz}$ and $D_0 \simeq 
10^{-8}\mathrm{cm}^{2}/\mathrm{sec}$. The estimate $(a_0/\xi_T)^2
\simeq 10^{-2}$ suggests only minor disclination motion during an
undulation period and the spectrum is unaffected for $0<l<6$.

\section{Dynamics and instabilities in cylindrical geometries}
\label{sec:6}

Next, we will discuss the influence of hexatic order on the
Plateau-Rayleigh instability \cite{chan,fabe} of cylindrical liquid
jets coated with a hexatic monolayer.  Because the Gaussian curvature
of a cylinder is zero, defect-free hexatic order is perfectly
compatible with this geometry in the absence of deformations. However,
the hexatic stiffness constant will resist deformations leading to a
nonzero Gaussian curvature. The complication of defects in the ground
state is absent for hexatic order on cylinders.

The surface vector of a cylinder of undeformed radius $R_0$ is given by 
\begin{equation}
\label{eq:6.1}
  \vec{R}(x)=\left ( R_0\cos \varphi,R_0 \sin \varphi,z \right ),
\end{equation}
with $x=(\varphi,z)$. See again Appendix~\ref{app:1} for the
fundamental geometrical quantities of the cylinder  in terms of
$\vec{R}$. 

The free energy of liquid cylinders (jets)  is given by
Eq. (\ref{eq:3.1}) where now $dA=R_0 d \varphi dz$.
The displacement field can again be chosen to be purely normal, i.e.,  
\begin{equation}
  \vec{R}'=\vec{R}+R_0 \zeta \vec{N},
\end{equation}
where now $\vec{N}(\varphi,z)=(\cos \varphi,\sin \varphi,0)$. The
displacement field can be expanded in plane waves
\begin{equation}
  \zeta(\varphi,z,t)=\sum_{k,m} r_{km}(t) e^{ikz} e^{im \varphi},
\end{equation}
where $k=2 \pi n/L$, with $n=0,\pm 1,\pm 2,...$ for a cylinder with
length $L$ (we assume periodic boundary conditions along the axis of
the cylinder for simplicity) and 
\begin{equation}
  \sum_{k,m} \equiv \sum_{n=-\infty }^{\infty } \sum_{m=- \infty
    }^{\infty }. 
\end{equation}

We neglect the density outside the jet
($\rho_v \ll \rho_l$), and  make an ansatz for the velocity potential
inside in terms of cylindrical coordinates $(r,\varphi,z)$, 
\begin{equation}
  \Phi(r,z,\varphi,t)= \sum_{k,m} A_{km}^<(t) e^{ikz} e^{im \varphi}
  I_m(kr),
\end{equation}
where $I_m(kr)$ is the Bessel function of the first kind of imaginary
argument \cite{grad}. 
Eqs. (\ref{eq:3.2}) and
(\ref{eq:3.4}) remain valid and yield
\begin{equation}
  A_{km}^<(t)=R_0 \frac{\dot{r}_{km}(t)}{kI_m'(kR_0)},
\end{equation}
where $I_m'(x)\equiv dI_m(x)/dx$.

If the cylinder is deformed, its surface area and volume
change. Eq. (\ref{eq:a4.3}) of Appendix~\ref{app:4} implies
\begin{equation}
  A'-A=2 \pi R_0 L \left [ r_{00} + \frac{1}{2} \sum_{k,m} {}^{'}\left
  ( R_0^2 k^2 
      +m^2\right ) |r_{km}|^2 \right],
\end{equation}
and with Eq. (\ref{eq:a4.5}) one has
\begin{equation}
  V'-V=2 \pi R_0^2 L \left ( r_{00} + \frac{1}{2} \sum_{k,m} {}^{'} 
  |r_{km}|^2 \right), 
\end{equation}
where the $k=0$, $m=0$ term is excluded from the sum $\sum_{k,m}
{}^{'}$.

Upon choosing
\begin{equation}
  r_{00}=-\frac{1}{2} \sum_{k,m} {}^{'}|r_{km}|^2,
\end{equation}
the displacement field keeps the volume fixed (corresponding to an
incompressible liquid jet) and the difference in interfacial free
energy between deformed and undeformed cylinder becomes
\begin{equation}
  F_i'-F_i=\pi R_0 L  \sigma \sum_{k,m} {}^{'} \left (R_0^2k^2+m^2-1
  \right   ) |r_{km}| ^2. 
\end{equation}
Note that this energy difference vanishes for $k=0$, $m=\pm 1$
deformations, corresponding to a uniform sideways translation.

We next calculate the 
hexatic contribution to the free energy of the deformed
cylinder. Eq. (\ref{eq:a4.12}) with background Gaussian curvature
$\K=0$ leads to the Gaussian curvature of the deformed state, namely 
\begin{equation}
  \K'(\varphi,z)=\sum_{k,m} k^2 r_{km}e^{ikz} e^{im \varphi}.
\end{equation}
Eq. (\ref{eq:4}), together with the representation of the Green's
function of the Laplacian on 
the cylinder, 
\begin{equation}
\G(x,x')=-\frac{R_0}{2 \pi L} \sum_{k,m} {}^{'} \frac{e^{ikz}
    e^{im \varphi} e^{-ikz'} e^{-im \varphi'}}{k^2R_0^2+m^2},
\end{equation}
then yields for the hexatic free energy of the deformed cylinder
\begin{equation}
  F_h'=\frac{\pi L}{R_0} K_A  \sum_{k,m} |r_{km}| ^2
  \frac{k^4R_0^4}{k^2R_0^2 +m^2}. 
\end{equation}
Eq. (\ref{eq:3.2}) now leads to 
\begin{equation}
\rho _l R_0 \ddot{r}_{km}(t) \frac{I_m(kR_0)}{k
  I_m'(kR_0)}=p_{km}=-\frac{\delta F_d'\{r_{km}\}}{2 \pi R_0^2 L\delta
  r_{km}^*}, 
\end{equation}
where 
\begin{equation}
  \Delta p(z,\varphi)=\sum_{k,m} p_{km} e^{ikz} e^{im \varphi}. 
\end{equation}
Upon setting $r_{km}=r^0_{km}e^{-i \omega(k,m)t}$ one finds the
fluctuation 
spectrum of a liquid cylinder with surface hexatic order, namely
\begin{equation}
\label{eq:6.17}
  \omega^2(k,m)=\frac{\sigma}{\rho_l R_0^3} \frac{
    I'_m(kR_0)}{I_m(kR_0)}R_0k \left [R_0^2k^2+m^2-1+\frac{K_A}{\sigma 
      R_0^2}   \frac{k^4R_0^4}{k^2R_0^2 +m^2} \right ]. 
\end{equation}

When $K_A=0$, Eq. (\ref{eq:6.17}) shows $\omega^2<0$ for $m=0$ and 
$R_0^2k^2<1$. The cylinder thus becomes unstable if 
$L>2 \pi R_0$, leading to the well-known 
Plateau-Rayleigh instability \cite{chan,fabe}. However, for $K_A \neq 0$ the
stability of liquid cylinders is enhanced by hexatic order. The
'fastest growing' mode $k_f$, which maximizes $(-\omega^2)$, is now
given by  
\begin{equation}
\left. \frac{d}{dx}
\right |_{x=R_0k}
\frac{I'_m(x)}{I_m(x)}x \left (1-x^2-\frac{K_A}{\sigma 
      R_0^2} x^2\right )=0,
\end{equation}
where $x=kR_0$. 
Thus, close to the hexatic-to-liquid transition where $K_A/\sigma
R_0^2 \simeq 1$,  one has $R_0^2k_f^2 \simeq 0.25$
compared with $R_0^2k_f^2 \simeq 0.48$ for $K_A=0$.  Thus, the
characteristic wavelength $\lambda_f=2 \pi/k_f$ of the undulations of
the unstable cylinder is significantly stretched by the presence of
hexatic order. As we shall see in the next section, hexatic order has
a similar effect in stabilizing multielectron bubbles against
fission.

\section{Hexatic dynamics and fission in multielectron bubbles}
\label{sec:4}

Next, we will discuss 
multielectron bubbles in liquid $ ^4\mathrm{He}$. These bubbles can
undergo both a freezing 
transition and a shape instability. Hexatic order 
affects both the fluctuation spectrum and the 
instability threshold for fission.
The free energy of a multielectron bubble $F_b=F_d+F_c$ is that of a
droplet (cf. Eq.~(\ref{eq:3.1})) with an additional Coulomb
contribution, i.e., 
\begin{equation}
\label{eq:4.1}
    F_b =
\sigma \int dA
 + \frac{1}{2} K_A \int dA  \ D _in^j D ^i n_j+ \frac{1}{2
   \varepsilon} \int dA \int  dA' \ 
 \frac{\rho(x)\rho(x')}{|x-x'|}. 
\end{equation}
Here, $\rho(x)$ denotes the charge distribution on the surface and
$\varepsilon$ is the  dielectric constant of liquid $ ^4\mathrm{He}$
\cite{foot3}. 
In an equilibrium fluid, $\rho=eN/4 \pi R_0^2$ for a sphere with $N$
electrons.

The assumptions underlying this theoretical description are: (i) The
electrons are restricted to the two-dimensional manifold given by the
liquid-vapor interface. This is justified since density functional
calculations (cf. e.g. \cite{shik78} and \cite{salo81}) show that the
electrons form a thin layer of thickness $\delta \ll R_0$ on the
surface with $\delta \simeq 1-4$nm.  (ii) One can neglect the effects
on the shape of the charged bubbles arising from: (a) applied electric
fields which trap or hold the electrons; (b) the movement of the
bubbles in the system; and (c) gravity. Assumption (a) is justified
since typical external electric fields are of the order $E \simeq
3\mathrm{kV/cm}$ \cite{leid97}, while a charged sphere with radius
$R_0 \simeq 10 \mu\mathrm{m}$ and $N \simeq 10^7$ produces an electric
field approximately 300 times larger, $E_{sph}=\frac{eN}{2R_0^2}\simeq
1 \mathrm{MV/cm}.$ Similar arguments apply to (b) since typical drag
forces are much weaker than the Coulomb forces. Assumption (c) is
justified since $R_0$ is much smaller than the capillary length $l_c$
of liquid helium, $l_c \equiv \sqrt{2 \sigma/\rho_lg}\simeq 0.6$mm.
Finally, (iii) we assume that the electrons can be treated
classically, i.e.  corrections arising from quantum mechanics can be
ignored. For electrons on charged bubbles this is justified since at
the melting temperature quantum mechanical corrections become only
relevant at higher densities, $n \gtrsim 10^{12}\mathrm{cm^{-2}}$
\cite{leid97}.

Upon once again neglecting defects by setting $s_{lm} \equiv 0$,
$l>0$, the stationary solutions of the free energy (\ref{eq:4.1}) are
spherical bubbles.  To determine the equilibrium radius we use the
partition function of a noninteracting ideal gas to describe the vapor
phase.  The free energy of a spherical bubble is then given by
\begin{equation}
  F(V,T) =\min_{R} \left \{k_BTN_{He} \left ( \log \frac{N_{He}
        \lambda_T^3}{4 \pi R^3/3} -1\right)+ p_{ex}\frac{
      4\pi}{3}R^3+4 \pi \sigma R^2+\frac{N^2 e^2}{2R \varepsilon}  \right \} ,
\end{equation}
where $N_{He}$ is the number of helium atoms, $\lambda_T$ the
thermal wavelength and $p_{ex}$ the pressure outside of the
bubble. Minimization with respect to the first term only 
yields, of course, ideal gas behavior, i.e.,  
\begin{equation}
p_{in}V=N_{He}k_BT, 
\end{equation}
where $p_{in}$ is the pressure inside the bubble.  However, by taking
all contributions into account the equilibrium radius $R_0$ of a
charged multielectron bubble is determined by the Laplace equation
\begin{equation}
\label{eq:7}
p_{in}-p_{ex}=
2 \frac{\sigma}{R_0} - \frac{(eN)^2}{8 \pi   R_0^4 \varepsilon}.
\end{equation}
Thus, one obtains (for $p_{in} \approx p_{ex}$) as
typical length scale for $R_0$ the
classical Coulomb radius
\begin{equation}
R_{cl}^3=\frac{(eN)^2}{16 \pi \sigma \varepsilon}.  
\end{equation}

Within the approximations described above, the fluctuation spectrum of 
a multielectron bubble with hexatic order can be calculated. Compared
with the discussion of Sect.~\ref{sec:3} the only 
difference arises from the 
Coulomb contribution. 
As derived in Appendix~\ref{app:4.3} one has
for $\rho_v \ll \rho_l$ (thus neglecting the density inside the
bubble)
\begin{equation}
  \Delta p_{lm}=\frac{(eN)^2}{4 \pi R_0^4 \varepsilon} (l-1) r_{lm}. 
\end{equation}
By combining the last equation  with Eqs. (\ref{eq:3.52}),
(\ref{eq:3.53}) and (\ref{eq:2.31})
one now finds (for $l>0$) \cite{foot:4}
\begin{equation}
\label{eq:4.5}
  \omega^2  = \frac{\sigma}{\rho_{l}R_0^3}(l-1)(l+1)
 \left [
    (l+2)-4 \frac{R_{cl}^3}{R_0^3}+\frac{K_A}{\sigma R_0^2}
    \frac{(l-1)(l+2)^2}{l(l+1)} \right ]. 
\end{equation}              
For $K_A=0$
spherical bubbles become unstable to fission if $R_0<R_{cl}$, i.e. 
$\omega^2(l=l_c)<0$ for $R_0<R_{cl}$ and
$l_c=2$ \cite{salo81,land}. 

\begin{figure}
  \begin{center}
    \mbox{\epsfxsize=8cm    \epsffile{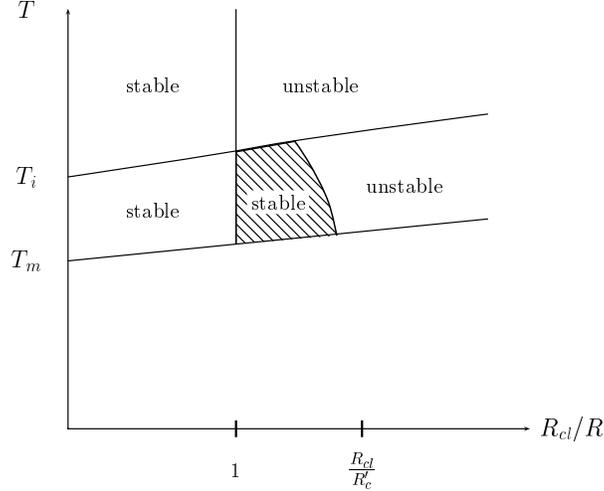}}
  \end{center}
\caption{\label{fig:4} Schematic stability  diagram for multielectron
  bubbles with hexatic order. The stability of a spherical bubble with
  radius $R$ and charge $eN$ depends on the temperature $T$. For
  hexatics in the temperature range $T_m \lesssim T \lesssim T_i$,
  bubbles with radii $R_{cl}<R<R_c'$ are stable whereas for $T \gtrsim
  T_i$ they are not.  The enhanced stability is due internal hexatic
  order. We expect an even larger region of enhanced stability for
  droplets with crystalline order \cite{lenz03}.}
\end{figure}

For $K_A \neq 0$ the 
stability of charged bubbles is {\em enhanced}
by the hexatic order of the electrons on the
sphere.  Thus, for $T_m<T<T_i$ the unstable mode is still  $l_c=2$,
but now 
$\omega^2(l_c=2)<0$ for $R_0<R_c'$ with $R_{c}'<R_{cl}$,
cf. Fig. \ref{fig:4}. 
The icosahedral 
symmetry of the deformed shape with  $s_{lm}\neq 0$ is too high to
have an influence on the fission instability which occurs at $l=2$,
justifying our neglect of disclination defects.

Because the electrons (which determine $K_A$) are far apart relative
to the helium atoms (which determine $\sigma$) $K_A/\sigma R_{cl}^2$
will be smaller than for droplets of supercooled liquid metals when
$R_0 \simeq R_{cl}$.  For helium-bubbles with $N \simeq 10^6$ one has
$R_{cl} \simeq 10 \mu$m and we expect (see Sect. \ref{sec:7}) that
$\frac{K_A}{\sigma R_{cl}^2}\simeq 10^{-4}$.  {\em Charged} metal
droplets also undergo the same fission instability.  Thus, it might be
possible to detect the onset of hexatic order by investigating the
stability of charged liquid droplets in Paul traps \cite{davi97}.

In the flat space limit ($l \rightarrow \infty$, $k=l/R_0$ fixed) one
obtains, with $n=eN/4 \pi R_0^2$,
\begin{equation}
\label{eq:4.7}
  \rho_{ex} \omega^2(k)=
  \sigma k^3-4 \pi k^2 \frac{n^2}{\varepsilon},
\end{equation}
in agreement with \cite{leid97}. As Eq. (\ref{eq:4.7}) shows, planar
He-interfaces are unstable to deformations with $k<k_c=4 \pi
n^2/\sigma \varepsilon$. This is the 'dimple' instability discussed in
the Introduction which triggers the initial creation of multielectron
bubbles.

\section{Dynamics in hexatic membranes with a spherical topology}
\label{sec:2}

As a last application we
discuss the influence of hexatic order on the fluctuations of
a spherical vesicle. The free energy $F$ of a hexatic membrane is
given by \cite{park92,nels87}
\begin{eqnarray}
    F & = & F_b+F_g+F_h \nonumber \\
& \equiv &   \frac{1}{2} \kappa \int dA \  (2H)^2+\kappa_G\int dA \ \K
 + \frac{1}{2} K_A \int dA  \ D _in^j D ^i n_j,
\label{eq:2.1}
\end{eqnarray}
where we have neglected spontaneous curvature, and $\kappa$ and
$\kappa_G$ are the mean and Gaussian rigidity, respectively.

The equilibrium shape of enclosed vesicles (provided fluids inside and
outside cannot equilibrate on experimental times scales) often
corresponds to the minimum of the free energy (\ref{eq:2.1}) with
prescribed surface area $A$ and volume $V$ \cite{schn84} (see also
Ref. \cite{miln87}). The area and volume constraints can be
implemented by considering the modified free energy
\begin{equation}
  \tilde{F}=F+F_a+F_{vol} \equiv F+ \int dA \sigma
  (x) + \int dV p(x).
\label{eq:2.3}
\end{equation}
Often, a constant pressure difference between out- and inside $p
\equiv p_{ex}-p_{in} \equiv p^>-p^<$ and surface tension $\sigma$ are
taken as Lagrange multipliers to enforce these constraints. Here, we
allow for both spatially varying surface tension $\sigma(x)$ and
pressure $p(x)$ to enforce local incompressibility (cf. Eq.
(\ref{eq:2.22}) below). In many situations, the average values of
$\sigma$ and $p$ will depend on the volume captured by a vesicle of a
given area at the moment of its formation.

In the following analysis we only consider surface shapes
which are topologically equivalent to a sphere. Then, the second term
of Eq.~(\ref{eq:2.1}) can be neglected, since it is a topological
invariant. We again temporarily ignore discrete disclination defects
and show later that their inclusion does not affect characteristic
frequencies $\omega(l)$ with $l<6$. 

Because $\kappa$ plays a similar role for vesicles as $\sigma$ plays
for droplets, hexatic order should lead here to similar effects on
the spectrum $\omega(l)$.  
For vesicles the dynamical fluctuations take place at very small
Reynolds numbers \cite{foot:2.1}. Therefore, the convective and the
inertial terms can be neglected in the Navier-Stokes equation 
(\ref{eq:2.14}) for the surrounding bulk fluid. Thus, here we must
solve the Stokes equation
\begin{equation}
\label{eq:2.15}
  \nabla p= \eta \nabla ^2 \vec{v},
\end{equation}
where  $\eta$ and $\vec{v}$ denote now the shear viscosity and the
velocity of the bulk fluid, assumed to be the same inside and outside
the membrane. 
The undulating membrane imposes the 
boundary condition
\begin{equation}
\left.  \vec{N} \cdot \vec{v}(\vec{r}) \right
|_{\vec{r}=\vec{R}_0(1+\zeta)}  = R_0 \dot{\zeta} \label{eq31},
\end{equation}
where $\vec{R}_0$ is given by Eq. (\ref{eq:2.2}) and the membrane
velocity by 
\begin{equation}
  R_0 \dot{\zeta} \equiv R_0 \frac{\partial
    \zeta(\theta,\varphi,t)}{\partial t }. 
\end{equation}
The following analysis simplifies somewhat by imposing boundary
conditions for the components 
$\nabla (\vec{N}\cdot \vec{v})$ and $\nabla \times \vec{v}$ instead of
$v_{\theta}$ and $v_{\varphi}$. Then, 
\begin{eqnarray}
\left. \vec{N} \cdot \nabla [\vec{N} \cdot \vec{v}(\vec{r})]
\right |_{\vec{r}=\vec{R}_0(1+\zeta)} & = &- \nabla \cdot (R_0
\dot{\zeta} \vec{N}), \label{eq32}\\
\left. \vec{N} \cdot [\nabla  \times \vec{v}(\vec{r})]
\right |_{\vec{r}=\vec{R}_0(1+\zeta)} & = & 0 \label{eq33},
\end{eqnarray}
where Eq. (\ref{eq32}) follows from the condition $\div \vec{v}=0$.

We proceed now as follows \cite{schn84}: For a given (deformed)
vesicle shape the bending forces on its surface are known. These
forces must be balanced by the viscous stresses. The induced flow
field $\vec{v}$ can be calculated by using Lamb's solution with these
boundary conditions.  Details about this solution have been given by
several authors, cf. e.g. Refs. \cite{schn84}, \cite{seif98}, and
\cite{happ}.  The main formulas are summarized in
Appendix~\ref{app:3}.

As shown in Appendix~\ref{app:3}, the viscous force 
$\vec{\Pi} \equiv \Pi_n \vec{N} + \vec{\Pi}_t$   associated
with Lamb's solution (see Eqs. (\ref{eq:a3.1}) and (\ref{eq:a3.4})) is
given by Eqs. (\ref{eq:C.12}), (\ref{eq:C.13}) and (\ref{eq:C.14}).
By using Eqs. (\ref{eq:a3.9}) and (\ref{eq:a3.10}), one then obtains
for the normal and tangential force differences 
\begin{equation}
\label{eq:19}
  \Pi_n^<(x)-\Pi_n^>(x)  =  \sum_{l=1}^{\infty } \sum_{m=-l}^{l}
  \dot{r}_{lm}(t) 
  \eta \frac{(2l+1)(2l^2+2l-3)}{l(l+1)} \Ylm(x),
\end{equation}
and
\begin{equation}
  \vec{\Pi}_t^<(x)-\vec{\Pi}_t^>(x)  =  \sum_{l=1}^{\infty }
  \sum_{m=-l}^{l} \dot{r}_{lm}(t) 
  R_0  \eta  \frac{2l+1}{l(l+1)} \nabla  \Ylm(x),
\end{equation}
where $\vec{\Pi}_t^<$ and $\vec{\Pi}_t^>$ refer to forces inside and
outside of the vesicle respectively. 

The tangential motion of the fluid along the membrane induces lipid
flow within the membrane itself. However, the membrane stays locally
incompressible. This constraint is enforced by the local surface
tension in Eq. (\ref{eq:2.3}), which balances the tangential component
of the stress vector. Then, with
\begin{equation}
\label{eq:2.21}
  \sigma(x,t)=\sigma_{0}+\sum_{l=1}^{\infty } \sum_{m=-l}^{l}
  \sigma_{lm}(t) \Ylm(x),
\end{equation}
one finds (for $l \geq 1$)
\begin{equation}
\label{eq:2.22}
  \sigma_{lm}(t)=\dot{r}_{lm}(t) R_0 \eta \frac{2l+1}{l(l+1)}.
\end{equation}

Since any non-stationary membrane configuration exerts a local force
on the surrounding fluid the shape displacement leads to a generalized 
pressure discontinuity $\Delta p(x)$ which balances 
the normal component of the 
pressure discontinuity between the inside and outside 
\begin{equation}
\label{eq:2.23}
\Pi_n^<(x)-\Pi_n^>(x)=\Delta p(x),
\end{equation}
where $\Delta p(x)$ can be expanded in spherical harmonics as in Eq.
(\ref{eq:2.23b}).  In order to determine the coefficients $p_{lm}$ in
Eq.  (\ref{eq:2.23b}) now the variation of the vesicle free energy has
to be calculated. Details are presented in Appendix~\ref{app:4}.

The deformed surface is characterized by 
the Gaussian
curvature (see Eq. (\ref{eq:a4.13}))
\begin{equation}
\label{eq:2.24}
\K'(x)  = \frac{1}{R_0^2}+\frac{1}{R_0^2} \sum_{l,m}(l-1)(l+2)
r_{lm}\Ylm(x),  
\end{equation}
and the mean curvature given by Eq. (\ref{eq:2.25}). 
The volume change is given by Eq. (\ref{eq:3.53p})
and the change in surface area by Eq. (\ref{eq:a4.6}), which becomes
\begin{equation}
  A'-A=R_0^2 \left ( 2\sqrt{4 \pi} r_{00}+\sum_{l=1}^{\infty }
    \sum_{m=-l}^{l} |r_{lm}|^2(1+l(l+1)/2) \right ) +
  \mathcal{O}(r_{lm}^3). 
\end{equation}
The volume constraint can again be implemented by considering 
displacements which fulfill Eq. (\ref{eq:2.29}). Then, upon defining
the excess area (relative to a sphere) by $\delta A=A-4 \pi R_0^2$,
the area constraint reads
\begin{equation}
\label{eq:2.30a}
\frac{1}{2}  R_0 ^2 \sum_{l=1}^{\infty }
    \sum_{m=-l}^{l} |r_{lm}|^2 (l-1)(l+2)= \delta A \equiv
    \mbox{const}. 
\end{equation}
Similarly, the area term of Eq. (\ref{eq:2.3}) for the deformed sphere
becomes
\begin{equation}
  F_a'=F_a+2 R_0^2\sum_{l,m} \sigma_{lm} r_{lm}^*
+ \frac{1}{2} \sigma_0  R_0 ^2 \sum_{l=1}^{\infty }
    \sum_{m=-l}^{l} |r_{lm}|^2 (l-1)(l+2). \label{eq:2.30}
\end{equation}
Finally, it has been shown in \cite{pete85}
and \cite{mors} that the bending energy of a quasi-spherical vesicle
is given by
\begin{equation}
  F_b'=F_b+\frac{1}{2} \kappa R_0^2 \int dA \zeta \mathcal{L} \zeta,
\end{equation}
where $F_b=8 \pi \kappa$ and
$\mathcal{L}$ can be evaluated on an undeformed sphere to leading
order in $\zeta$,
\begin{equation}
  \mathcal{L}=D^iD_iD^jD_j+\frac{2}{R_0^2}D^iD_i. 
\end{equation}
Upon using the decomposition (\ref{eq:10}) of $\zeta(x,t)$, 
one finds
\begin{equation}
\label{eq:2.32}
  F_b'=F_b+ \frac{1}{2} \kappa \sum_{l=1}^{\infty } \sum_{m=-l}^{l}
|r_{lm}|^2 l(l+1)(l-1)(l+2).
\end{equation}

We now insert Eqs. (\ref{eq:2.30}), (\ref{eq:2.31}), and
(\ref{eq:2.32}) into Eq. (\ref{eq:2.23b}) (with $\tilde{F}'$ replacing
$F_d'$) and set again  $r_{lm}=r^0_{lm}(t)e^{-i \omega(l)t}$. 
Upon equating the pressure discontinuity in (\ref{eq:2.23b}) to
the normal force difference (\ref{eq:19})
we find for the
fluctuation spectrum of spherical vesicles with hexatic order
(for $l>0$)
\begin{equation}
  \omega(l) = -i\frac{\Gamma(l)}{\eta R_0^3 } (l-1)(l+2)
\nonumber \left [\kappa 
  l(l+1)+ \sigma_0 R_0^2+K_A \frac{(l-1)(l+2)}{l(l+1)}\right ],
\label{eq:2.34}
\end{equation}
with 
\begin{equation}
  \Gamma(l) \equiv \frac{l(l+1)}{(2l+1)(2l^2+2l-1)}.
\end{equation}
Note that 
$\omega(l)$ vanishes for $l=1$,  corresponding to 
translations of the vesicle as a whole.

The eigenfrequencies $\omega(l)$ explicitly depend on the $l=0$
component of the tension $\sigma_0$ which acts as Lagrange-multiplier
for the area.  Since the value of $\sigma_0$ is generally not known,
one has to use the area constraint (\ref{eq:2.30a}) (and the
fluctuation-dissipation theorem) to express $\sigma_0$ in terms of
$\delta A$.  Which modes pick up the excess area depends on the ratio
$\tilde{\gamma}=\sigma R_0^2/\kappa$.  As shown in Ref. \cite{miln87}, for
floppy membranes (i.e. small $\tilde{\gamma}$) each mode contributes
equally to the excess area, whereas for stiff membranes (i.e. large
$\tilde{\gamma}$) only the lowest mode ($l=2$) will develop a large
amplitude \cite{foot:5.2}.

For the purpose of estimating the effect of hexatic order, we proceed
in a different way.  The area constraint becomes much easier to handle
if the volume constraint can be neglected. By assuming that the vesicle is
permeable to both water and larger molecules, the area constraint can
be incorporated directly (again to leading order in the $r_{lm}$'s) by
choosing
\begin{equation}
r_{00}=- \frac{1}{2\sqrt{4 \pi}}\sum_{l=1}^{\infty }
    \sum_{m=-l}^{l} |r_{lm}|^2\left(1+\frac{l(l+1)}{2} \right). 
\end{equation}
Then, $A'=A$ and one can set $\sigma_0=0$ in Eq. 
(\ref{eq:2.34}).  Equivalently, we focus on floppy vesicles, formed
under conditions such that $\sigma R_0^2 \ll \kappa$. 

In the flat space limit of $\sigma_0=0$, large $R_0$ and $l \gg 1$
with $k \equiv l/R_0$ fixed, one has
\begin{equation}
\label{eq:2.38}
  \omega \simeq -i \frac{1}{4\eta} \left [ \kappa k^3+
    \frac{K_A}{R_0^2}k\right ] .
\end{equation}
As for the liquid droplets discussed in Sect. \ref{sec:3}, the hexatic
contribution drops out as $R_0 \rightarrow \infty$ and we recover the
result for undulation modes of a flat fluid bilayer \cite{seif97}.
The frequency shift (\ref{eq:2.34}) depends on the ratio $K_A/\kappa$.
However, for large vesicles we expect that $K_A \simeq 4\kappa$ (a
{\em universal} result for flat hexatic membranes at long wavelengths
\cite{davi87}) leading to a frequency enhancement in Eq.
(\ref{eq:2.34}) (with $\sigma_0=0$) by a factor $\simeq 13/9 \simeq
1.44$ for the $l=2$ quadrupole mode.
Note the relatively pronounced effect of hexatic order. In general, we 
expect the largest change in the characteristic frequencies for {\em
  floppy} vesicles with $\tilde{\gamma}=\sigma R_0^2/\kappa \ll 1$.

Similar to droplets, 
the presence of an icosahedral array of defects in hexatic membranes
leads to an equilibrium configuration with a deformed surface.  
For vesicles, the 
complete first variation of the free energy $\tilde{F}$ (see
Eq. (\ref{eq:2.3})) reads
\begin{eqnarray}
  \delta^{(1)} \tilde{F}& = & \frac{1}{2}\kappa \int d^2x \sqrt{g(x)}
  \left (  
(2H(x))^2  \delta^{(1)} g(x)+8H(x)  \delta^{(1)}H(x) \right )  \nonumber \\
& &  + \sigma_0 \int d^2x 
\sqrt{g(x)}   \delta^{(1)} g(x) + \int d^2x 
\sqrt{g(x)} p  \delta^{(1)} V(x)\nonumber \\ & & 
 - K_A \int d^2 x \sqrt{g(x)} \frac{1}{R_0^2}
  \sum_{l=1}^{\infty } \sum_{m=-l}^{l}
 \frac{(l-1)(l+2)}{l(l+1)} s_{lm}\Ylm(x) \sum_{l',m'}r^*_{l'm'}
{\mathrm{Y}_{l'm'}^*}(x),
\end{eqnarray}
where (again) $\delta^{(1)}g$ is given by Eq. (\ref{eq:a4.3}),
$\delta^{(1)}H$ by Eq. (\ref{eq:a4.15}) and $\delta^{(1)}V$ by
Eq. (\ref{eq:a4.5}). Since integration by parts shows that 
\begin{equation}
  \int d^2x \sqrt{g(x)} H(x) \nabla^2 \zeta(x)=  \int d^2x \sqrt{g(x)}
  \zeta(x) \nabla^2 H(x), 
\end{equation}
the equilibrium shape equation for quasi-spherical vesicles with
hexatic order becomes now
\begin{eqnarray}
\lefteqn{  p+2 \sigma_0 H(x) -4H(x) \kappa (H^2(x)-\K(x)) -2 \kappa
\nabla^2 H(x)=} \nonumber \\
& &  K_A  \frac{1}{R_0^3}
  \sum_{l=1}^{\infty } \sum_{m=-l}^{l}
 \frac{(l-1)(l+2)}{l(l+1)}\Ylm (x)s_{lm}.
\end{eqnarray}
Upon making again the ansatz (\ref{eq:2.37}) 
(where $2H_0 \sigma_0 +p=0$)
and using 
$\K=\K_0+2 \widetilde{\delta H}/R_0$, one finds 
for the mean curvature
\begin{equation}
  2H(x)=2H_0+ \frac{1}{R_0}
  \sum_{l=1}^{\infty } \sum_{m=-l}^{l}
\frac{K_A}{\kappa l (l+1)+\sigma_0 R_0^2}
  \frac{(l-1)(l+2)}{l(l+1)}s_{lm}\Ylm(x)  , 
\end{equation}
where the
static surface deformation coefficients in the
ground state are given by 
\begin{equation}
 r_{lm}^0=s_{lm}\frac{K_A}{\kappa l^2(l+1)^2 +\sigma_0 R_0^2 l(l+1)}
\end{equation}
for $l>0$ and $r_{lm}^0=0$ for $l=m=0$. 

However, icosahedral symmetry again insures that $s_{lm}=0$, and hence
$r_{lm}^0=0$, unless $l=6,10,12,...$, at least for floppy,
approximately spherical vesicles. Hence, corrections of order $s_{lm}$
have no influence on the frequencies $\omega(l)$ for small $l$.  Just
as for hexatics on liquid droplets, disclination motion is negligible
during an undulation period and the dispersion relation
(\ref{eq:2.34}) remains valid for $0<l<6$. This can be seen by using
an estimate similar to that of Sect.~\ref{sec:2.2}. Here, $\omega(l=2)
\simeq 40$Hz for a $1 \mu$ vesicle and the disclination diffusion
constant is $D_5 \approx (a_0/\xi_T)^2D_{lipid}$, where $D_{lipid}
\simeq 10^{-8}\mathrm{cm}^{2}/\mathrm{sec}$ and $(a_0/\xi_T)^2 \simeq
10^{-2}$.

Finally, we show that our assumption that  the regular part of the
bond-angle field  relaxes
rapidly on the time scale of undulation modes is indeed justified for
vesicles with hexatic order. As follows from Eq. (\ref{eq:9.8}) the
relaxational  dynamics for 
$\theta^{reg}$ is described by
\begin{equation}
\eta_m \frac{d \theta^{reg}}{d t}=K_A \nabla^2 \theta^{reg},  
\end{equation}
where $\eta_m$ is the shear surface viscosity of the membrane. 
For simplicity, we have neglected in the last equation
defects which are immobile on the time scale of ripples anyhow. 
Upon expanding  $\theta^{reg}$
in terms of spherical harmonics 
\begin{equation}
 \theta^{reg}=\sum_{l=0}^{\infty } \sum_{m=-l}^{l}
 \theta^{reg}_{lm}(t)\Ylm(x), 
\end{equation}
and setting 
\begin{equation}
 \theta^{reg}_{lm}=\theta^0_{lm}e^{-i \omega_{\theta}(l)t} , 
\end{equation}
one finds
\begin{equation}
\omega_{\theta}(l)=-i \frac{K_A}{\eta_m R_0^2} l(l+1). 
\end{equation}
Thus,  by comparing with the undulation spectrum $\omega$ given by
Eq. (\ref{eq:2.34})
one obtains with $\eta_m=\eta h$ (where $h \simeq 100-1000 \mathrm{nm}$
for bilayers) for floppy vesicles and large $l$
\begin{equation}
\frac{\omega(l)}{\omega_{\theta}(l)} \simeq \frac{\kappa}{4K_A}
\frac{h}{R_0} l.   
\end{equation}
Thus, $\omega(l) \ll \omega_{\theta}(l)$ for $h \ll R_0$, justifying
our assumption for large vesicles.  Similar arguments apply to
spherical droplets and bubbles and cylindrical geometries.

\section{Experimental consequences}

\label{sec:7}

We conclude by discussing the implications of our results for possible
experiments.  As discussed in the Introduction, one of our main
motivations was to search for dynamic signatures of hexatic order in
curved geometries, order which is difficult to detect by other means.
As shown in Sect.~\ref{sec:2}, the frequency shift
due to hexatic order is particularly large for vesicles. The $l=2$
quadrupole relaxation rate can be enhanced by a factor $\simeq 1.44$
due to bond-orientational order and
should be experimentally detectable by measuring 
the decay rate of  fluctuations by, e.g., video or
fluorescence microscopy \cite{schn84}. 

For liquid droplets the influence of hexatic order on
the fluctuation spectrum is somewhat weaker. For the magnitude 
of the hexatic stiffness one has the estimate \cite{halp78}
\begin{equation}
\label{eq:11}
K_A =C(T)  \left (\frac{\xi_T}{a_0} \right )^2,
\end{equation}
where $a_0$ is the lattice constant, $\xi_T$ is a translational
correlation length which diverges as $T$ approaches $T_m$ and $C(T)$
is a prefactor.  When $T \gg T_m$ one has $C(T) \simeq 2 E_c
(a/a_0)^2$ \cite{halp78}, where $E_c$ is the energy of a dislocation
core with diameter $a \approx a_0$.  On the other hand, for the liquid
metal droplets $\sigma \simeq k_BT/a_0^2$ and with $C(T) \simeq k_BT$
one has $K_A/\sigma R_0^2 \simeq 1$ for $\xi_T \simeq R_0$. Thus,
hexatic order should also have experimentally relevant consequences
for the fluctuation spectrum of liquid droplets and cylinders when the
translational correlation length grows to become comparable to the
sphere size or cylinder radius. As we saw in Sect.~\ref{sec:6}, the
Plateau-Rayleigh instability will also be modified for $K_A \neq 0$.
As the liquid cylinder decays into a chain of droplets the distance
between the droplets will depend on the ratio $K_A/\sigma R_0^2$, as
follows from Eq. (\ref{eq:6.17}). For $K_A/\sigma R_0^2 \simeq 1$ one
has $\lambda_f=2 \pi/k_f \simeq 12.6R_0$ as typical distance between
the droplets in contrast to $\lambda_f \simeq 9.0R_0 $ appropriate to
$K_A=0$ \cite{chan,fabe}.

For multielectron bubbles $\sigma$ is determined by the distance $b_0$
between the helium atoms.  Here, we expect that $\sigma \simeq 3
k_BT_m/b_0^2$, where $T_m$ is the melting temperature and $b_0 \simeq
3 $\AA \ for liquid helium. From the experiments by Grimes and Adams
\cite{grim79} it is known that for the planar case the dimensionless
ratio $\Gamma_m \equiv \frac{e^2 \sqrt{\pi n}}{k_BT_m} \simeq 140$ and
$T_m=0.73\mathrm{K}$ for a charge density $n=10^9\mathrm{cm^{-2}}$.
Thus, for $ ^4\mathrm{He}$ one has $\sigma\simeq 3 \cdot
10^{-4}\mathrm{J/m^2}$ at $T \approx T_m$. For helium bubbles with $N
\simeq 10^6$ one has $a_0=(4 \pi R_{cl}^2 /N)^{1/2} \simeq 300$\AA \ and
the critical radius $R_{cl} \simeq 10 \mu$m. By using the $T=0$ result
$E_c \simeq 0.1 e^2/a_0 $ \cite{fish79} one finds $K_A/\sigma R_{cl}^2
\simeq 10^{-4}$.  It is worth mentioning that $K_A/\sigma R_{cl}^2 \sim
1/\sqrt{N}$ since $K_A \sim NE_c$ and $E_c \sim R_{cl}^2/N^{3/2}$. Thus,
the influence of hexatic order will be somewhat stronger for bubbles
with smaller $N$.

Thus, hexatic order has only a weak influence on the stability of
multielectron bubbles. But with sophisticated experimental setups it
still might be possible to measure its effects.  A possible scenario
to study fission could, e.g., make use of the fact that bubbles become
compactified as they move under the influence of an external field
away from the liquid-vapor interface towards the positive electrode
immersed in the liquid.  Thus, if for $T>T_i$ multielectron bubbles
with an initial radius $R_0$ are stable to fission up to a distance
$h$ from the interface then they will be observed at a somewhat larger
distance $h+\delta h$ for $T_m<T<T_i$.

It also might be possible to experimentally confirm our theoretical
predictions in the regime of stable bubbles by measuring the
characteristic frequencies of an oscillating bubble which is
stabilized by the balance of bouancy and electric field forces at a
constant height below the surface.  Our predictions also apply to
multiion bubbles which experimentally can be realized by charging
helium films with ions instead of electrons \cite{leid95}.

Finally, for charged droplets in Paul traps similar effects should be
observable by, e.g., trapping a droplet and shooting charges on it
until it undergoes fission. For this purpose it might be necessary to
use more conventional liquids and to perform the experiments at much
higher temperatures. Fission with ordered ions on droplets can still
be achieved by using ions with high charge $Ze$ since $T_m \propto Z^2
e^2 \sqrt{n}$.  For example, for droplets with $R_0 \simeq 10 \mu$m
and $N \simeq 10^6$ the melting transition should occur at room
temperature for $Z=7$. Provided the surface tension of the liquid is
large enough the droplets will still be stable at these parameter
values.

In summary, we have shown that two-dimensional hexatic order should
lead to experimentally observable effects in a variety of systems with
spherical and cylindrical geometry. Similar effects could also occur
in related systems such as, e.g., cylindrical vesicles. Here, we
expect that the laser-induced pearling instability will be modified by
the presence of hexatic order \cite{barz94,nels95}.

Crystalline order will also alter the fluctuation spectrum of
spherical and cylindrical droplets and membranes
\cite{zhan93,komu92,erdi00}.  The effects discussed here should be
even {\em larger} if the hexatic phase is bypassed and one freezes
directly into a two-dimensional solid with shear modulus $\mu$. The
resulting frequency shifts can be estimated by replacing $K_A$ by $\mu
R_0^2$ in the formulas above. Details will be given in an upcoming
publication \cite{lenz03}.

We thank S. Balibar, G. Gabrielse, N. Goddard, J. Lidmar and R. Pindak
for helpful discussions.  This work has been supported by the NSF
through Grant No.~DMR97-14725 and through the Harvard MRSEC via Grant
No.~DMR98-09363. P.~L. acknowledges support by the Deutsche
Forschungsgemeinschaft through an Emmy-Noether fellowship (Le
1214/1-1).

\newpage

\begin{appendix}

\section{Parameterization of shape}
\label{app:1}
\setcounter{equation}{0}
\renewcommand{\theequation}{A\arabic{equation}}

Here, we review the elementary differential geometry needed to
determine all relevant geometrical properties, i.e. the first and
second fundamental forms, of a surface in terms of its surface vector
$\vec{R}$. For a more extensive review, see \cite{davi}. For
simplicity, we specialize here to spherical and cylindrical surfaces.
{\em Deformations} of these shapes are treated in
Appendix~\ref{app:4}.

For an undeformed sphere with constant radius $R_0$ one can choose 
$x=(x^1,x^2) \equiv(\theta,\varphi)$ 
with polar coordinates $\theta$ and $\varphi$. The 
surface vector is then given  by 
\begin{equation}
\label{eq:a1}
\vec{R}(\theta,\varphi) =R_0 \left( \sin \theta
  \cos \varphi, \sin \theta 
  \sin \varphi, \cos \theta\right).
\end{equation}

\begin{figure}
  \begin{center}
    \begin{tabular}{cc}
\mbox{\epsfxsize=6cm \epsffile{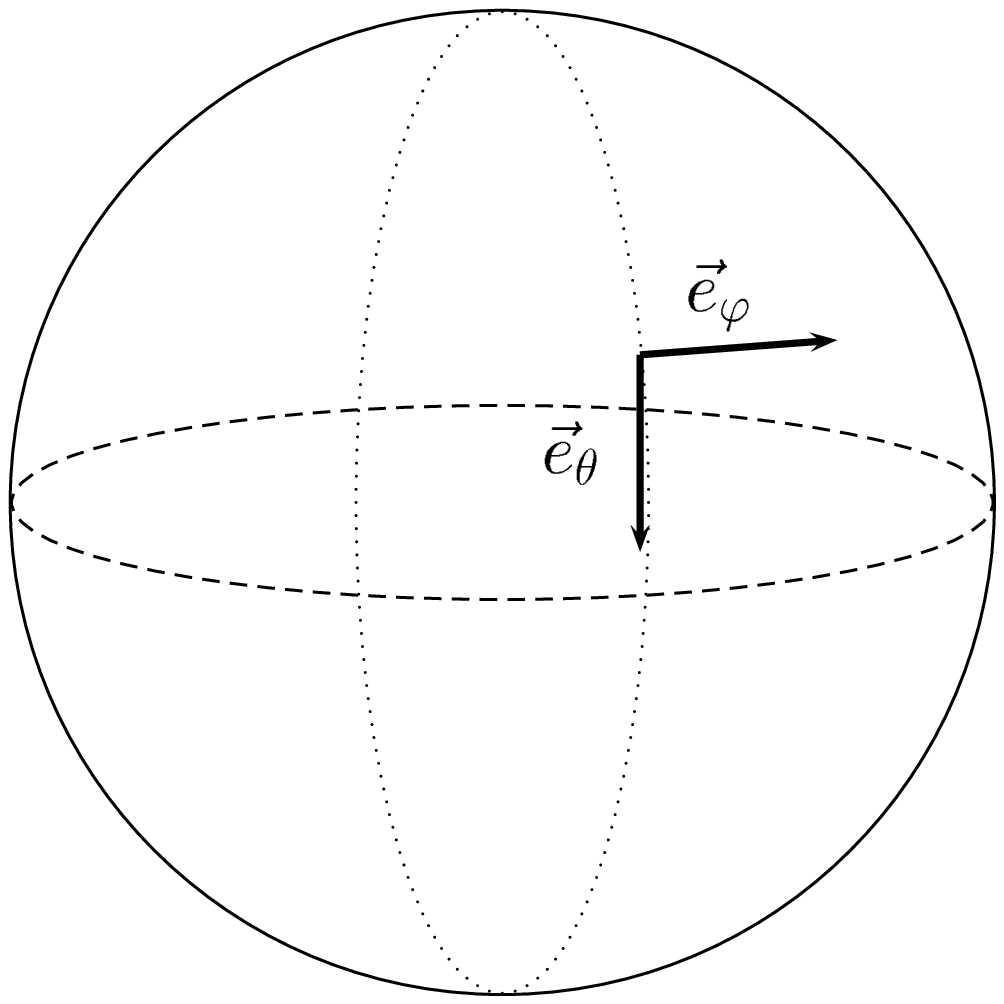}} &
\begin{minipage}{6cm}
\vspace*{-5cm}
\mbox{\epsfxsize=7cm \epsffile{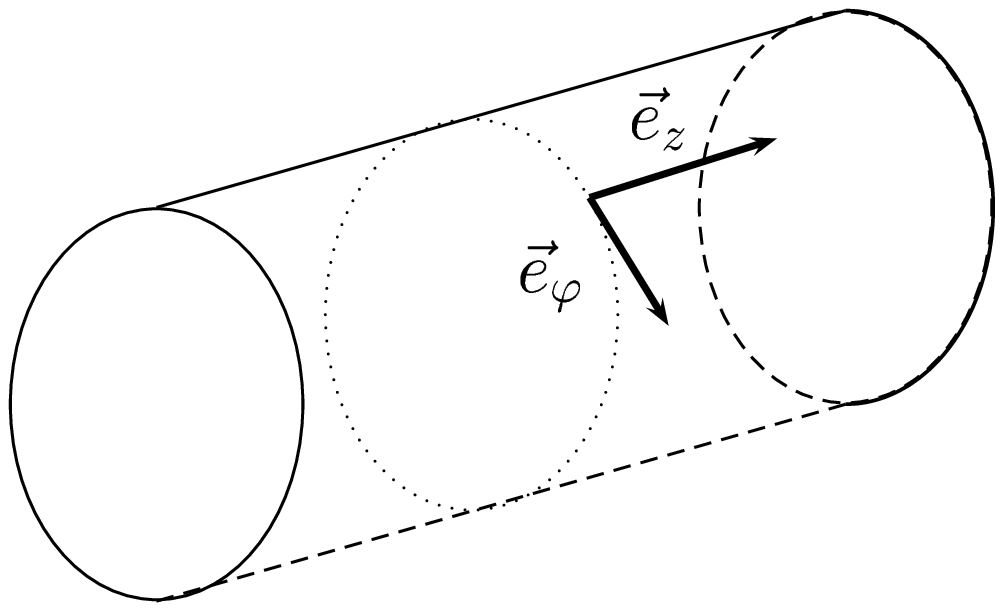}}
\end{minipage}
    \end{tabular}
    \caption{\label{fig:5} Orthonormal basis vectors $\vec{e}_{\theta}$ and
      $\vec{e}_{\varphi}$ of the tangential plane of the sphere and
      basis vectors $\vec{e}_{\varphi}$ and $\vec{e}_{z}$  of the
      tangential plane of the cylinder. One has
      $\vec{e}_{\theta}=\vec{R}_1/R_0$,
      $\vec{e}_{\varphi}=\vec{R}_2/R_0\sin \theta$ for the sphere, see Eq.
      (\ref{eq:2.2}). For the cylinder,
      $\vec{e}_{\varphi}=\vec{R}_1/R_0$ and $\vec{e}_{z}=\vec{R}_2$,
      see Eq. (\ref{eq:6.1}).  }
  \end{center}
\end{figure}

The first fundamental form is defined by
\begin{equation}
\label{gl:3.1}
  g _{ij} \equiv \vec{R}_i \cdot \vec{R}_j,
\end{equation}
with the covariant  vectors
\begin{equation}
 \vec{R}_j \equiv \vec{R}_{,j} \equiv \partial _{s^j} \vec{R} \equiv
 \frac{\partial \vec{R}}{\partial s^j}  . 
\end{equation}
On the sphere the vectors 
$\{\vec{R}_1(x),\vec{R}_2(x)\}$ 
form an orthogonal (but not orthonormal) basis in the tangential plane at 
the point $x$, cf. Fig. \ref{fig:5}.
The  contravariant  components are given by 
$\vec{R}^k=g^{ki}\vec{R}_i$, where $g^{ij}$ is the inverse of
$g_{ij}$, i.e. 
$g^{ij}g_{jk}=\delta^i_k$. We use the summation convention throughout.
With the parameterization of Eq. (\ref{eq:a1}) we have
\begin{equation}
  \vec{R}_1=R_0 \left (\cos \theta \cos \varphi,\cos \theta \sin
    \varphi, -\sin \theta \right ), \quad
  \vec{R}_2=R_0 \left (-\sin \theta \sin \varphi,\sin \theta \cos 
    \varphi, 0 \right). 
\end{equation}

The area element is generally given by $dA=\sqrt{g}dx^1dx^2$, where 
\begin{equation}
\label{gl:3.2}
  g \equiv \det (g _{ij}).   
\end{equation}
For a sphere the first fundamental form is given by
\begin{equation}
(g_{ij})=
  \left ( 
    \begin{array}{cc}
R_0^2 & 0 \\ 0 & R_0^2 \sin^2 \theta
    \end{array}
\right )
\end{equation}
and $dA=R_0^2 \sin\theta d\theta d \varphi$.  

\begin{figure}
  \begin{center}
    \begin{tabular}{cc}
\mbox{\epsfxsize=6cm \epsffile{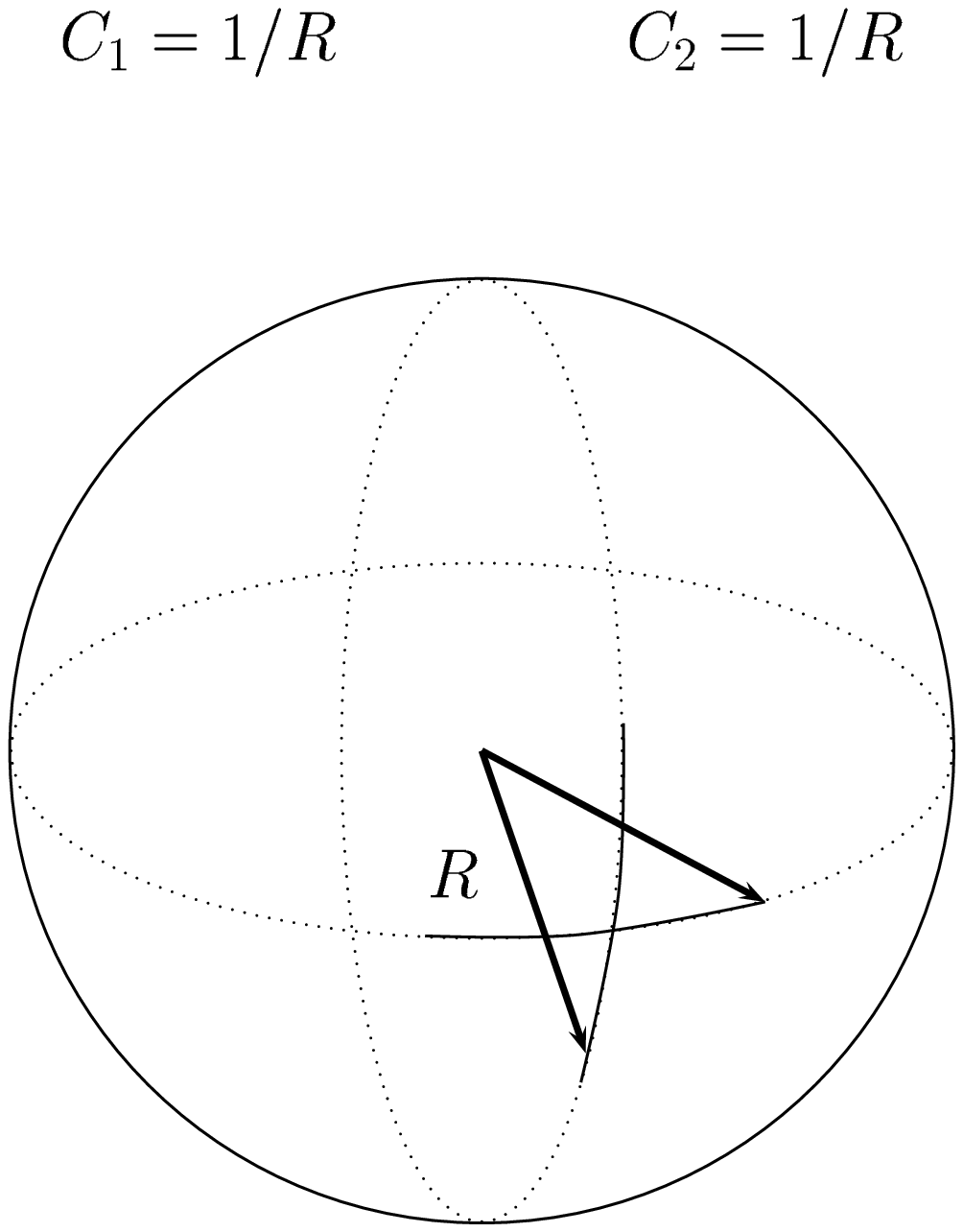}} &
\begin{minipage}{6cm}
\vspace*{-5cm}
\mbox{\epsfxsize=7cm \epsffile{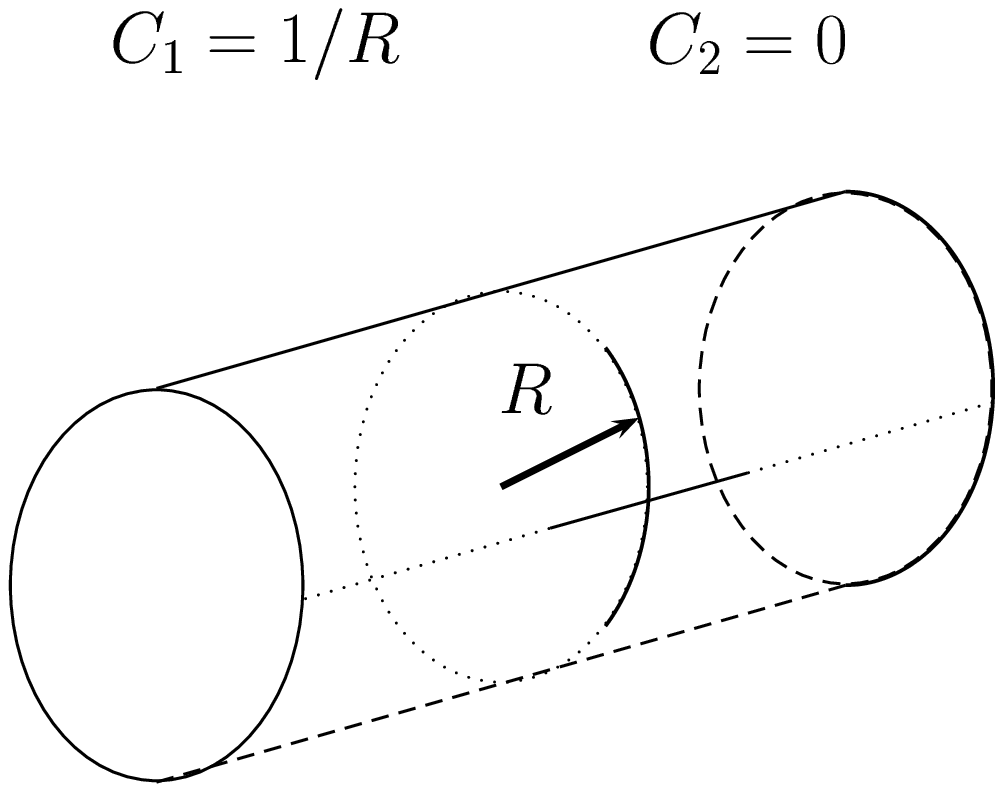}}
\end{minipage}
    \end{tabular}
    \caption{\label{fig:6} Principal curvatures for spheres and
      cylinders. In both cases the principal directions of curvature
      are drawn. For a sphere both principal curvatures are identical, 
      i.e. $C_1=C_2=1/R$. For a cylinder one has 
      $C_1=1/R$, $C_2=0$. The principal curvatures are the
      eigenvalues of $(-b_i^j)$. Thus,
      $2H=-(C_1+C_2)$ and $\K=C_1C_2$. }
  \end{center}
\end{figure}

The mean curvature $\H$ and the
Gaussian curvature $\K$ are determined by the {\em second} fundamental form,
which is defined by
\begin{equation}
\label{gl:3.8}
  b _{ij} \equiv \vec{R} _{ij} \cdot \vec{N}=-\vec{R}_i \cdot \vec{N}_j,
\end{equation}
where $\vec{N}$ is the  (local) unit normal vector to the surface
\begin{equation}
\vec{N} (x^1,x^2)\equiv \frac{\vec{R}_1(x^1,x^2) \times 
\vec{R}_2(x^1,x^2)}{\sqrt{g}},
\end{equation}
and the second equality of Eq. (\ref{gl:3.8}) follows from 
$\vec{R}_i \cdot \vec{N}=0$. In terms of the second fundamental form,
and upon raising an index of $b_{ij}$ via the operation 
$g^{ik}b_{kj} \equiv b^i_j$, we have
\begin{equation}
\label{gl:3.9}
  2\H=-b_i^i  \quad \mbox{ and }  \quad \K= \det (b^i_j)=\frac{b}{g}
  \equiv 
  \frac{\det (b_{ij})}{\det (g_{ij})}.  
\end{equation}
See  Fig. \ref{fig:6}. 
For a sphere one has
\begin{equation}
  \vec{N}=\frac{\vec{R}}{R_0}=\left( \sin \theta \cos \varphi, \sin
    \theta   \sin \varphi, \cos \theta\right)
\end{equation}
and the second fundamental form becomes
\begin{equation}
(b_{ij})=
  \left ( 
    \begin{array}{cc}
-R_0 & 0 \\ 0 & -R_0 \sin^2 \theta
    \end{array}
\right ).
\end{equation}
It is easy to check that $b^i_j=-\delta^i_j/R_0$ and hence, $\H=1/R_0$
and $\K=1/R_0^2$ in this case.  

The covariant derivative of a vector with covariant components $a_i$
and contravariant components $a^j$ is defined by \cite{krey}
\begin{equation}
  \D_i a_j=a_{j,i}-a_k \Gamma^k_{ij} \quad \mbox{ and } \quad 
  \D_i a^j=a^j_{,i}+a^k \Gamma^j_{ki}. 
\end{equation}
Here, the $\Gamma_{ij}^k$ are the Christoffel symbols of the 
second kind. They are 
related to the Christoffel symbols of the first kind $\Gamma_{ikj}$ by 
$\Gamma^k _{ij}=g^{kl}\Gamma_{ilj}$, where 
\begin{equation}
\label{gl:3.11}
\Gamma_{ikj} \equiv \frac{1}{2} \left( \partial_i g_{kj}+\partial_j
  g_{ik}-\partial_k g_{ij}\right).
\end{equation}
Note that the $\Gamma_{ikj}$ are symmetric in the first and last
index, i.e.,  
\begin{equation}
\label{gl:3.12}
  \Gamma_{ikj}=\Gamma_{jki}.
\end{equation}

\begin{figure}
  \begin{center}
    \mbox{\epsfxsize=8cm    \epsffile{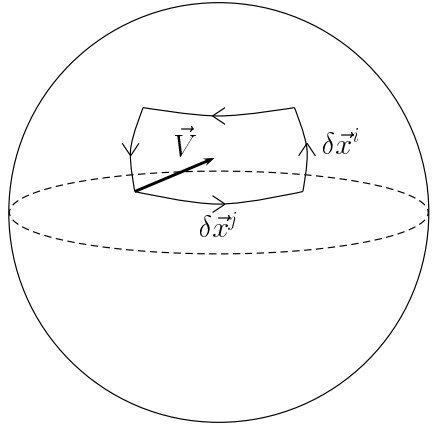}}
  \end{center}
\caption{\label{fig:7}
Geometrical interpretation of the Riemann curvature tensor on the
sphere. Parallel 
transport of the vector $\vec{V}$ along the path shown in the figure
leads to a rotation 
$(\vec{V}'-\vec{V})^l=R_{kji}^lV^k \delta x^j \delta x^i$, where
$R_{kji}^l$ is the Riemann curvature tensor, defined in
Eq. (\ref{gl:3.19}). As a consequence, on a curved surface covariant
derivatives do not commute. Indeed, $[D_i,D_j]\eta^i=\K \eta_j$, where 
$\K$ is the Gaussian curvature which in two dimensions determines all
components of $R_{kji}^l$. }
\end{figure}

For a sphere with polar coordinates we have
\begin{equation}
  \label{eq:3.26}
  \Gamma_{\theta\theta}^{\theta}=\Gamma_{\theta\theta}^{\varphi}=
  \Gamma_{\theta\varphi}^{\theta}=\Gamma_{\varphi\varphi}^{\varphi}=0, 
\Gamma_{\theta\varphi}^{\varphi}=\frac{\cos \theta}{\sin
  \theta},\Gamma_{\varphi\varphi}^{\theta}=-\sin \theta \cos \theta. 
\end{equation}
As can be shown by a direct calculation, the covariant derivatives of
$g_{ij}$, $g^{ij}$, and $\delta^i_j$  vanish.
Covariant derivatives do not commute in general. 
For scalars one has 
$\D_i\D_jf=\D_j\D_if$  if 
$\partial _i \partial _jf=\partial _j \partial _if$.
However, for vectors the commutator of covariant derivatives 
is given by
\begin{equation}
\label{gl:3.18}
  [\D_i,\D_j]a_k=a_lR^l_{kji}.
\end{equation}
Here,  $R^l_{kji}$ is the (mixed) Riemann curvature tensor
\begin{equation}
\label{gl:3.19}
  R^l_{kji} \equiv \partial_{j} \Gamma^l_{ki} - \partial_{i}
  \Gamma^l_{kj} + 
  \Gamma^m_{ki} \Gamma^l_{mj} - \Gamma^m_{kj} \Gamma^l_{mi}.
\end{equation}
In two spatial dimensions one has the particularly simple relation 
\begin{equation}
\label{eq:3.22}
  R^l_{kji}=g_{km}\gamma^{lm}\gamma_{ji}\K,
\end{equation}
where $\gamma^{ij}$ is the antisymmetric contravariant tensor,
\begin{equation}
\label{gl:3.3}
  \gamma^{ij} \equiv (\delta^i_1 \delta^j_2-\delta_2^i
  \delta_1^j)/\sqrt{g} ,
\end{equation}
with covariant counterpart
$ \gamma_{kl}=\gamma^{ij}g_{ik}g_{jl}$. Note, 
Eq. (\ref{eq:3.22}) shows that the Gaussian curvature is determined by 
the first fundamental form only. This is the content of the 
famous theorema egregium of Gauss. 
A geometrical interpretation of the Riemann curvature tensor is given
in Fig. \ref{fig:7}.

For a sphere the nonzero components of the  Riemann curvature tensor
are 
\begin{equation}
\label{eq:3.28}
  R^1_{212}=-R^1_{221}=\frac{b}{g_{11}}=\sin^2 \theta, \quad 
  -R^2_{112}=R^2_{121}=\frac{b}{g_{22}}=1.
\end{equation}

Covariant derivatives have another useful property: integrals over
covariant derivatives of vector fields can be integrated by parts,
similar to vector fields in flat space. 
This is the content of the divergence theorem: for an arbitrary closed
surface $M$ (i.e. a surface with no boundary, $\partial M=0$) one has
\begin{equation}
\label{eq:div} \label{eq:3.23}
  \int_{M} dA \D_i v^i=0
\end{equation}
for any vector $v^i$. Here 
$\D_i v^i=\frac{1}{\sqrt{g}}\partial _i(\sqrt{g})v^i) \equiv \div v$
is the divergence of $v$ on a curved surface.
Finally, the Laplacian is defined by $\Delta=\D^i \D_i$. It becomes on 
the sphere
\begin{equation}
  \label{eq:3.27}
\Delta_{sph}=\frac{1}{R_0^2} \left \{    \frac{1}{\sin^2
    \theta} \frac{\partial ^2 }{\partial
      \varphi^2}+ \frac{1}{\sin \theta} \frac{\partial }{\partial
      \theta} \left (\sin \theta \frac{\partial }{\partial \theta}
    \right )     \right \}.
\end{equation}
Note that the eigenfunctions of $\Delta_{sph}$ are spherical
harmonics 
\begin{equation}
  \Delta_{sph} \Ylm=\frac{-l(l+1)}{R_0^2}\Ylm.
\end{equation}

For undeformed cylinders with radius $R_0$ 
one can choose $x=(\varphi,z)$. 
The surface vector is then given by
\begin{equation}
  \vec{R}(x)=\left ( R_0\cos \varphi,R_0 \sin \varphi,z \right ). 
\end{equation}
Here, the first fundamental form is given by
\begin{equation}
(g_{ij})=
  \left ( 
    \begin{array}{cc}
R_0^2 & 0 \\ 0 & 1
    \end{array}
\right )
\end{equation}
and $dA=R_0  d \varphi dz $.  

The second fundamental form becomes
\begin{equation}
(b_{ij})=
  \left ( 
    \begin{array}{cc}
-R_0 & 0 \\ 0 & 0 
    \end{array}
\right ).
\end{equation}
Thus,  $\K_0=0$ and  $H_0=1/2R_0$. 

Since for cylinders $g_{ij}$ is independent of $\varphi$ and $z$ 
all Christoffel symbols vanish, i.e.
$\Gamma_{ij}^k=0$. Thus, $\partial_i=\D_i$ and the Laplacian becomes
for this geometry
\begin{equation}
  \Delta_{cyl}=\partial ^i \partial_i=\frac{1}{R_0^2} \frac{\partial
  ^2}{\partial \varphi ^2}+ \frac{\partial ^2}{\partial z^2}. 
\end{equation}
The eigenfunctions are plane waves 
\begin{equation}
\Delta_{cyl}  e^{ikz} e^{im \varphi} = -\frac{1}{R_0^2} (k^2 R_0^2 +
m^2)e^{ikz} e^{im \varphi}.
\end{equation}

\section{Hexatic free energy}
\label{app:2}
\setcounter{equation}{0}
\renewcommand{\theequation}{B\arabic{equation}}

In this Appendix we derive a useful representation of the hexatic free
energy for vesicles, droplets and multielectron bubbles. The
discussion is kept rather general since it has to be valid for both
deformed and undeformed spheres and cylinders.  The discussion here
generalizes earlier approaches \cite{nels87,davi87} to (arbitrary)
curved geometries with defects. See also Ref. \cite{bowi00}.

Consider, 
the free energy $F_h$
introduced
in Sect.~\ref{sec:3}, i.e., 
\begin{equation}
\label{eq:9.1}
  F_h=\frac{1}{2}K_A \int d^2 x \sqrt{g} \D_i n^j\D^in_j. 
\end{equation}
We now introduce an ortho{\em normal} basis $\{\vec{e}_{\alpha}\}$ of
the tangential space. (Note that orthonormality is not guaranteed for
the tangent basis vectors discussed in Appendix~\ref{app:1}; for example,
the "natural" basis vectors $\vec{R}_{\theta}$ and $\vec{R}_{\varphi}$ for
the sphere are merely orthogonal). Then, for an arbitrary vector field
$\vec{V}$ one has the contravariant components
\begin{equation}
  \vec{V}=V^{\alpha} \vec{e}_{\alpha}, \quad 
  \vec{e}_{\alpha} \cdot 
  \vec{e}_{\beta} =\delta_{\alpha \beta}. 
\end{equation}
Here, Greek indices have been used to distinguish this basis from the 
$\{\vec{R}_i\}$. The covariant derivative of $\vec{V}$ is now given in
terms of
\begin{equation}
  \D_i V_{\alpha}=\vec{e}_{\alpha} \cdot (\partial _i \vec{V}),
\end{equation}
so that
\begin{equation}
  \D_i V_{\alpha}=\partial _i V_{\alpha}- \omega_{i \beta
    \alpha} V^{\beta}, 
\end{equation}
with $V_{\alpha}=V^{\beta}\vec{e}_{\beta} \cdot \vec{e}_{\alpha}$ and
\begin{equation}
\label{eq:9.5}
  \omega_{i \beta \alpha}=\vec{e}_{\beta} \cdot \partial _i
  \vec{e}_{\alpha} 
\end{equation}
is the  affine connection  appropriate for this basis \cite{naka}. 
For orthonormal $\vec{e}_{\alpha}$ one has
$\omega_{i \beta \alpha}=-\omega_{i \alpha \beta}$. Thus, one can 
define a  covariant vector field  $A_i$ by
\begin{equation}
 \omega_{i \alpha \beta}=\varepsilon_{\alpha \beta}A_i, 
\end{equation}
where
$\varepsilon_{\alpha \beta} \equiv\delta_{\alpha}^1
\delta_{\beta}^2-\delta^2_{\alpha}  \delta^1_{\beta}$ is the 
Levi-Civita symbol.

We now set, 
$\vec{n}=n^{\alpha}\vec{e}_{\alpha}$. Since
$n^in_i=n^{\alpha}n_{\alpha}=1$ it is possible to introduce 
an bond-angle field $\Theta$ as angle between
$\vec{n}$ and the  local reference 
frame $\{\vec{e}_{\alpha}\}$with
\begin{equation}
  n_{\alpha=1}=\cos \Theta, \quad n_{\alpha=2}=\sin \Theta.
\end{equation}
Then, 
$\partial_in_{\alpha}=\varepsilon_{\alpha \beta}n^{\beta}\partial
_i\Theta$, 
and it is easy to show that Eq. (\ref{eq:9.1}) becomes 
\begin{equation}
\label{eq:9.8}
  F_h=\frac{1}{2}K_A \int d^2x \sqrt{g(x)} \left (\D^i \Theta+A^i
  \right ) \left (\D_i \Theta+A_i \right ). 
\end{equation}
For six-fold bond-oriented order, we identify $\theta$ with 
$\theta+2 \pi/6$, which determines the minimum charge of disclination
defects. 
The vector field $A_i$ is a "vector potential" associated with the Gaussian
curvature, i.e.,  
\begin{equation}
\label{eq:9.9}
  \D_iA_j-\D_jA_i=\partial _iA_j-\partial _jA_i=-\K (x)
  \gamma_{ij},
\end{equation}
see Ref. \cite{davi}. The last result
is a consequence of Eq. (\ref{eq:3.22}) and of the representation 
of  the curvature tensor in this basis, namely
\begin{equation}
  R^{\alpha}_{\beta ji} = \partial_j \omega^{\alpha}_{\beta i}
-\partial _i\omega^{\alpha}_{\beta j}+
  \omega^{\gamma}_{\beta i}\omega^{\alpha}_{\gamma j}
  -\omega^{\gamma}_{\beta j}\omega^{\alpha}_{\gamma i}.
\end{equation}
Note that Eq. (\ref{eq:9.9}) allows an explicit construction for the
vector potential for an arbitrary closed surface with a given
Gaussian curvature, that is
\begin{equation}
\label{eq:9.10}
  A_j(x)=-g^{kl} \gamma_{lj}D_k(x) \int d^2 x' \sqrt{g(x')}
  \G_g(x,x') \K(x').
\end{equation}
Here, $\G_g(x,x')$ is the Green's function for the Laplacian $\Delta_g$,
\begin{equation}
\label{eq:b12}
\Delta_{g}  \G_{g}(x,x') 
 =\frac{\delta(x-x')}{\sqrt{g}}. 
\end{equation}
Note that the Laplacian $\Delta_{g}$ depends explicitly on the  metric
$g$, since  
\begin{equation}
  \Delta_{g} f  \equiv \D^i\D_i f= \frac{1}{\sqrt{g}} \partial _i\left
    (\sqrt{g}     \partial ^i f \right ). 
\end{equation}

We now divide 
the field $\Theta$  into a regular part
$\Theta^{reg}$ and a singular part $\Theta^{sing}$ which represents
the contribution of defects. The regular part
$\Theta^{reg}$ fulfills $\D_i\D_j \Theta^{reg}=\D_j\D_i\Theta^{reg}$;
we assume that this part relaxes rapidly on the time scale of shape
deformations. The singular part $\Theta^{sing}$ is related to 
the disclination  density $s(x)$ by
(cf. Eq. (\ref{eq:2.5})) 
\begin{equation}
  \gamma^{ij}\D_i \D_j \Theta^{sing}=s(x) \equiv 
\frac{1}{ \sqrt{g}} \sum_{i} q_i \delta(x-x_i),
\end{equation}
where $q_i=\pm 2 \pi/6$ for disclinations in a hexatic. 
Thus,
\begin{equation}
\label{eq:9.11}
  \D_j(x) \Theta^{sing}(x)=g^{kl} \gamma_{lj}D_k(x) \int d^2 x' \sqrt{g(x')}
  \G_g(x,x') s(x'),
\end{equation}
where $\G_g(x,x')$ is the same Green's function as in
Eq. (\ref{eq:9.10}). 
Note that in the gauge defined by Eq. (\ref{eq:9.10}), one has 
\begin{equation}
  \D_i \D^i \Theta^{sing} +\D_i A^i =0,
\end{equation}
which is the Euler-Lagrange equation of the functional (\ref{eq:9.8}). 

By using Eqs. (\ref{eq:9.10}) and (\ref{eq:9.11}) the hexatic
free energy (\ref{eq:9.8}) becomes for an arbitrary manifold with
metric $g$ and Gaussian curvature $\K(x)$
\begin{equation}
\label{eq:9.15}
    F_h  =  -\frac{1}{2} K_A \int d^2x  \sqrt{g(x)} \int d^2 x' 
\sqrt{g(x')} \left (\K(x)-s(x) \right)
 \G_{g}(x,x') \left (\K(x')-s(x') \right) ,
\end{equation}
where we have integrated by parts and used Eq. (\ref{eq:b12}). 

In the planar case, the ground state  has no defects,
i.e. $s(x)=0$. Then, using 
$\G_g(x,x')=\left (\frac{1}{\Delta} \right )_{x,x'}$,
Eq. (\ref{eq:9.15}) reduces to the  Liouville 
action 
\begin{equation}
    F_h  =  -\frac{1}{2} K_A \int d^2x  \sqrt{g(x)} \int d^2 x' 
\sqrt{g(x')}  \K(x) \left (\frac{1}{\Delta_{xx'}} \right )
\K(x') .
\end{equation}

\section{Free energy of  deformed spheres and cylinders}
\label{app:4}
\setcounter{equation}{0}
\renewcommand{\theequation}{C\arabic{equation}}

As the spherical shape gets displaced its free energy changes.  In
this Appendix we calculate the corresponding contributions arising
from (i) the interfacial free energy and the bending energy
(Sect.~\ref{app:4.1}), (ii) the hexatic free energy
(Sect.~\ref{app:4.2}), and (iii) the Coulomb energy
(Sect.~\ref{app:4.3}). We begin by discussing the geometrical
properties associated with shape deformations.

\subsection{Geometrical properties and variations of the mean and
Gaussian curvature}
\label{app:4.1}

The displacement of spherical shapes parameterized by Eq.
(\ref{eq:2.8}) leads to new tangent vectors
\begin{eqnarray}
  \vec{R}_i' & = & \vec{R}_i -R_0 \zeta b_i^k \vec{R}_k + R_0 \zeta_i
  \vec{N} \label{eq:a4.1} \\
& = &  \vec{R}_i(1+\zeta) + R_0 \zeta_i \vec{N} .  
\end{eqnarray}
The first fundamental form then changes according to 
\begin{equation}
  \delta g _{ij} \equiv g_{ij}'-g_{ij}
 =  \vec{R}'_i \cdot
  \vec{R}'_j-\vec{R}_i\cdot \vec{R}_j= -2 R_0 \zeta b_{ij} -R_0
  \zeta^2 b_{ij} + R_0^2
  \zeta_i \zeta_j. 
\end{equation}
Correspondingly, the change in the area element is given by  
\begin{eqnarray}
  \sqrt{g+\delta g}
& = &  \sqrt{g} \left (1+2H R_0 \zeta +\K R_0^2 \zeta^2
    +\frac{1}{2} R_0^2 
    \zeta^i \zeta_i \right ) +\mathcal{O}(\zeta^3) \label{eq:a4.3}\\
& = &
\sqrt{g} \left (1+2 \zeta + \zeta^2
    +\frac{1}{2} R_0^2
    \zeta^i \zeta_i \right ) +\mathcal{O}(\zeta^3) \label{eq:a4.4} . 
\end{eqnarray}

The  volume change is given by
\begin{equation}
  \delta V= \int dA R_0\left ( \zeta+HR_0\zeta^2 \right
  )+\mathcal{O}(\zeta^3) \label{eq:a4.5}, 
\end{equation}
while the area of the deformed sphere is 
\begin{equation}
\label{eq:a4.6}
  A'=\int d^2x \sqrt{g+\delta g}. 
\end{equation}
The spherical harmonic decomposition (\ref{eq:10}) together with Eqs.
(\ref{eq:a4.4}) and (\ref{eq:2.29}) then leads to Eq. (\ref{eq:3.53}).

The change in the second fundamental form has only to be known up to 
first order in $\zeta$. The  normal vector 
of the deformed sphere is given by
\begin{equation}
\vec{  N}'=\vec{N}- R_0\zeta^i \vec{R}_i + \mathcal{O}(\zeta^2).
\end{equation}
Therefore, the second fundamental form changes according to
\begin{equation}
\label{eq:30}
  (b_{ij})'=\vec{R}_{ij}' \cdot \vec{N}'=b_{ij}+R_0\D_j
  \zeta_i+\zeta b_{ij}+\mathcal{O}(\zeta^2). 
\end{equation}
With 
\begin{equation}
  (g^{jk})'=g^{jk}+2 R_0 \zeta b^{jk} +
  \mathcal{O}(\zeta^2),
\end{equation}
one finds
\begin{equation}
(b_i^j)'=-\frac{1}{R_0}\delta_i^j(1-\zeta)+R_0\D^j \zeta_i +
\mathcal{O}(\zeta^2).  
\end{equation}
The resulting Gaussian curvature of the deformed surface is given by
\begin{equation}
  \K'=\det(b_i^j{}')=\frac{b'}{g'}=\frac{\det(b'_{ij})}{g+\delta g}+
  \mathcal{O}(\zeta^2)  .
\end{equation}
Therefore,
\begin{eqnarray}
\K'(x)-\K(x)  & = & 
 -2R_0H\K \zeta + R_0\gamma^{ik}
  \gamma^{jl}b_{ij}\D_l
  \zeta_k + \mathcal{O}(\zeta^2) \label{eq:a4.12}\\
& = &  -2\K \zeta-\nabla^2 \zeta+\mathcal{O}(\zeta^2) ,
\label{eq:a4.13}
\end{eqnarray}
which leads, upon expanding $\K(x)$ and $\zeta(x)$ in spherical
harmonics, to Eq. (\ref{eq:2.24}).  

Similarly, the mean curvature of the deformed surface is given by
\begin{equation}
\label{eq:a4.14}
  -2 H'=(b_i^i)'=(g^{ij})'(b_{ij})'=(g^{ij}+\delta
  g^{ij})(b_{ij}+\delta 
  b_{ij}) + \mathcal{O}(\zeta^2).
\end{equation}
Therefore, 
\begin{equation}
  2H'=2H-2R_0 \zeta (2 H^2-\K)-R_0\D^i \zeta_i
+\mathcal{O}(\zeta^2), \label{eq:a4.15} 
\end{equation}
which leads via spherical harmonics to Eq. (\ref{eq:2.25}). 

Although not derived here, Eqs. (\ref{eq:a4.1}), (\ref{eq:a4.3}),
(\ref{eq:a4.5}), (\ref{eq:a4.12}) and (\ref{eq:a4.15}) actually hold
for {\em general } geometries with mean curvature $H(x)$ and Gaussian
curvature $\K(x)$. These formulas will be needed in the analysis of
Sects. \ref{sec:2.2}, \ref{sec:6} and \ref{sec:2}.

\subsection{Hexatic free energy}
\label{app:4.2}

Next, we  calculate the change in
hexatic free energy. Since Eq. (\ref{eq:9.15}) holds for arbitrary
manifolds one has for the deformed sphere
\begin{eqnarray}
    F_h'  & = &  -\frac{1}{2} K_A \int d^2x  \sqrt{g+\delta g} \int d^2 x' 
\sqrt{g+\delta g} \nonumber \\
& & \times \left (\K'(x)-s'(x) \right)
\G_{g+\delta g}(x,x') \left (\K'(x')-s'(x') \right) . \label{eq:a4.17}
\end{eqnarray}
Here, $\G_{g+\delta g}(x,x')$ is the inverse Laplacian
defined on the deformed sphere, i.e., 
\begin{equation}
\Delta_{g+\delta g}  \G_{g+\delta g}(x,x') 
 =\frac{\delta(x-x')}{\sqrt{g+\delta g}} .
\end{equation}
Furthermore, $\K'$ is given by Eq. (\ref{eq:a4.13}). 
As the surface gets displaced the position of the defects might
change. This can be taken into account by introducing a 
disturbed defect
distribution 
\begin{eqnarray}
  s'(x) & = & s(x)+\frac{1}{R_0^2} \sum_{l=0}^{\infty } \sum_{m=-l}^{l}
  \delta s_{lm} \Ylm(x) \nonumber\\
& = & \K_0 +\frac{1}{R_0^2}\left ( \delta s_{00} \mathrm{Y}_{00}(x) +
\sum_{l=1}^{\infty }   
[s_{lm}+\delta s_{lm}] \Ylm(x)\right ). 
\end{eqnarray}
The coefficient $\delta s_{00}$ is determined by disclination ''charge
conservation'', i.e., 
\begin{equation}
  \int d^2 x \sqrt{g+\delta g} s'(x)=4 \pi.
\end{equation}
Thus,
\begin{equation}
  \delta s_{00}=-2 r_{00}. 
\end{equation}
Since the coefficients $\delta s_{lm}$ are of order $\zeta$, $s'(x)$
and $\K'(x)$ 
differ only in order $\zeta$. Therefore, one can replace in
Eq. (\ref{eq:a4.17}) 
$\sqrt{g+\delta g}$ by $\sqrt{g}$ and $\G_{g+\delta g}(x,x')$ by
$\G_g(x,x')$. Since for the sphere \cite{bowi00}
\begin{equation}
  \G(x',x'')=-
  \sum_{l=1}^{\infty } \sum_{m=-l}^{l}
  \frac{\Ylm(\theta',\varphi')\Ylm^*(\theta'',\varphi'')}{l (l+1)} ,
\label{eq:86b}
\end{equation}
one finally gets
\begin{equation}
\label{eq:a4.23}
    F_h' =  \frac{1}{2}K_A \sum_{l=1}^{\infty }\sum_{m=-l}^{l}
\frac{\left |r_{lm}(l-1)(l+2) - (s_{lm}+\delta s_{lm})
\right |^2}{l(l+1)} +\mathcal{O}(\zeta^2
  s_{lm}),
\end{equation}
a result needed in Sects.~\ref{sec:3} and \ref{sec:2}. 
\subsection{Coulomb energy}
\label{app:4.3}

The force (per unit area) acting on a dielectric (with dielectric
constant $\varepsilon$) contributed by an arbitrary electric field
$\vec{E}$ is given by \cite{land}
\begin{equation}
  p=\frac{E^2}{\varepsilon 8 \pi}.
\end{equation}
Then, in terms of the electrostatic potential $\psi$, $\vec{E}=-
\nabla \psi$ with $\nabla^2 \psi=0$, and one can make 
the ansatz (for $r>R_0(1+\zeta)$)
\begin{equation}
  \psi= \frac{eN}{r}+\sum_{l,m} c_{lm} \Ylm \left ( \frac{R_0}{r}
  \right ) ^{l+1}.
\end{equation}
The liquid-vapor interface has to be a surface of constant
potential. Thus, 
$\psi(r=R_0+R_0 \zeta)=$const. Upon setting 
$c_{lm}=cr_{lm}$, where $c$ is a constant, one then finds
\begin{equation}
  c_{lm}=\frac{eN}{R_0}r_{lm}
\end{equation}
and 
\begin{equation}
  \vec{E}=-\vec{e}_{r} \left . \frac{\partial }{\partial r} \right
  |_{r=R_0+\zeta R_0} \psi(r)= \frac{eN}{R_0^2} \left ( 1+ \sum_{l,m}
    (l-1) r_{lm} \Ylm \right )\vec{e}_{r}.
\end{equation}
Thus, 
\begin{equation}
  p=\frac{(eN)^2}{8 \pi R_0^4 \varepsilon}\left (1+2 \sum_{l,m}
  (l-1)r_{lm} \Ylm 
+ {\mathcal O}(r_{lm}^2)
  \right ). 
\end{equation}

\section{Lamb's solution for spherical ripples}
\label{app:3}
\setcounter{equation}{0}
\renewcommand{\theequation}{D\arabic{equation}}

Here, we present the main features of 
Lamb's solution for obtaining the fluid velocity fields from the
surface stresses. We follow here closely the presentations in
\cite{schn84}, \cite{seif98}, and 
\cite{happ}, where more details can be found.  

The inner solution [i.e. for $r<R_0(1+\zeta)$] of the Stokes equation
(\ref{eq:2.15}) with viscosity $\eta$
is given by  
\begin{equation}
\label{eq:a3.1}
  \vec{v}^<=\sum_{l=1}^{\infty }\left ( \nabla
    \varphi_l^<+\frac{l+3}{2 \eta (l+1)(2l+3)} r^2 \nabla
    \P_l^<-\frac{l}{\eta (l+1)(2l+3)} \vec{r} \P_l^<\right ), 
\end{equation}
with the velocity potential function
\begin{equation}
  \varphi_l^<(r,x,t)=\sum_{m=-l}^{l} \varphi_{lm}^<(t) \Ylm(x) \left (
    \frac{r}{R_0}\right ) ^l,
\end{equation}
and  the hydrostatic pressure $p_<=\sum_{l}p_l^<(r,x,t)$ with 
\begin{equation}
  \P_l^<(r,x,t)=\sum_{m=-l}^{l} \P_{lm}^<(t) \Ylm(x) \left (
    \frac{r}{R_0}\right ) ^l.
\end{equation}

The outer solution can be obtained by performing the
replacement  
$l \rightarrow -(l+1)$ in the formulas above. Thus, 
\begin{equation}
\label{eq:a3.4}
  \vec{v}^>=\sum_{l=1}^{\infty }\left ( \nabla
    \varphi_l^>-\frac{l-2}{2 \eta l(2l-1)} r^2 \nabla
    \P_l^>+\frac{l+1}{\eta l(2l-1)} \vec{r} \P_l^>\right ). 
\end{equation}
Here, 
\begin{equation}
  \varphi_l^>(r,x,t)=\sum_{m=-l}^{l} \varphi_{lm}^>(t) \Ylm(x) \left (
    \frac{R_0}{r}\right ) ^{l+1}
\end{equation}
and
\begin{equation}
  \P_l^>(r,x,t)=\sum_{m=-l}^{l} \P_{lm}^>(t) \Ylm(x) \left (
    \frac{R_0}{r}\right ) ^{l+1}.
\end{equation}

The boundary conditions become
\begin{equation}
  \left.  \vec{N} \cdot \vec{v}(\vec{r}) \right
|_{\vec{r}=\vec{R}_0(1+\zeta)} =  \sum_{l,m}\left \{ \frac{l}{2
    \eta (2l+3)}R_0 \P_{lm}^<(t)+ \frac{l}{R_0}\varphi_{lm}^<(t)
\right \} 
\Ylm(x) 
\label{eq:a3.7}
\end{equation}
and
\begin{eqnarray}
\lefteqn{\left. \vec{N} \cdot \nabla [\vec{N} \cdot \vec{v}(\vec{r})]
-\nabla \cdot \vec{v}(\vec{r})
\right |_{\vec{r}=\vec{R}_0(1+\zeta)}  =}  \nonumber \\ 
& & \sum_{l,m}\left \{
  \frac{l(l+1)}{2 
    \eta (2l+3)} \P_{lm}^<(t)+\frac{1}{R_0^2}l(l-1) \varphi_{lm}^<(t)
\right \}  
\Ylm(x). 
\label{eq:a3.8}
\end{eqnarray}
By comparing Eqs. (\ref{eq:a3.7}) and (\ref{eq:a3.8}) with the 
right  hand sides of Eqs.~(\ref{eq31}) and (\ref{eq32}) one then
obtains 
\begin{equation}
  \P_{lm}^<  = -\dot{r}_{lm}\eta \frac{(2l+3)(l-1)}{l} , \quad
  \P_{lm}^>  = \dot{r}_{lm}\eta \frac{(2l-1)(l+2)}{l+1}
\label{eq:a3.9} 
\end{equation}
and
\begin{equation}
  \varphi_{lm}^<  =  \dot{r}_{lm}R_0^2 \frac{(l+1)}{2l}, \quad
  \varphi_{lm}^>  =  \dot{r}_{lm}R_0^2 \frac{l}{2(l+1)}.
\label{eq:a3.10} 
\end{equation}
The stress vector associated with this velocity field is given by 
\begin{eqnarray}
  \vec{\Pi} & = & \Pi_n \vec{N} + \vec{\Pi}_t \nonumber \\
& = & -\vec{N}p +\eta \left (\frac{\partial \vec{v}}{\partial
    r}-\frac{\vec{v}}{r} \right ) +\frac{\eta}{r} \nabla (\vec{r}
\cdot \vec{v}). 
\end{eqnarray}
Here, the inner normal component is given by
\begin{equation}
  \vec{\Pi}^<_n=   \vec{\Pi}^<_n (\vec{r}=(\vec{R}_0+\zeta
  \vec{R}_0)^-)= 
\sum_{l,m} \left \{ 2 \frac{\eta}{R_0^2} l (l-1)
    \varphi_{lm}^< + \frac{l^3-4l-3}{(l+1)(2l+3)} \P_{lm}^<\right \}
  \Ylm(x). 
\label{eq:C.12}
\end{equation}
The  outer normal component reads
\begin{equation}
  \vec{\Pi}^>_n=   \vec{\Pi}^<_n (\vec{r}=(\vec{R}_0+\zeta
  \vec{R}_0)^+)= 
\sum_{l,m} \left \{ 2 \frac{\eta}{R_0^2} (l+2) (l+1)
    \varphi_{lm}^> - \frac{l^3+3l^2-l}{l(2l-1)} \P_{lm}^>\right \}
  \Ylm(x) .
\label{eq:C.13}
\end{equation}
Finally, the difference between the inside and outside tangential
component is given by 
\begin{eqnarray}
  \vec{\Pi}_t^<-\vec{\Pi}_t^> & = & \sum_{l,m} 
\left \{
2 \frac{\eta}{R_0} \left ( (l-1)\varphi_{lm}^<+(l+2)
  \varphi_{lm}^>\right )
\right . \nonumber \\
& & + \left .
\frac{l(l+2)}{(l+1)(2l+3)} \P_{lm}^<-\frac{(l^2-1)}{l(2l-1)}\P_{lm}^>
 \right \} \nabla  \Ylm(x). 
\label{eq:C.14}
\end{eqnarray}

\end{appendix}

\newpage

\begin{appendix}

\end{appendix}


\begin{thebibliography}{99999}
\bibitem{halp78} D.~R. Nelson
  and B.~I. Halperin, Phys. Rev. B {\bf 19}, 2457 (1979).
\bibitem{kost73} J.~M. Kosterlitz and D.~J. Thouless, J. Phys. C {\bf
    6}, 1181 (1973); A.~P. Young, Phys. Rev. B {\bf 19}, 1855 (1979). 
\bibitem{land37} L.~D. Landau, Phys. Z. Sowjetunion {\bf 11}, 26
  (1937). 
\bibitem{grim79} C. C. Grimes and G. Adams, Phys. Rev. Lett. {\bf 42},
  795 (1979).
\bibitem{glat88} D. C. Glattli, E. Y. Andrei and F. I. B. Williams,
  Phys. Rev. Lett. {\bf 60}, 420 (1998) and
  references therein. 
\bibitem{devi84} G. Deville, A. Valdes, E. Y. Andrei and
  F. I. B. Williams, Phys. Rev. Lett. {\bf 53}, 588 (1984).
\bibitem{fish79} D. S. Fisher, B. I. Halperin and R. Morf,
  Phys. Rev. B {\bf 20}, 4692 (1979).
\bibitem{morf79} R. H. Morf, Phys. Rev. Lett. {\bf 43}, 931 (1979).
\bibitem{chou98} C. F. Chou, A. J. Jin, S. W. Hui, C. C. Huang and
  J. T. Ho, Science {\bf 280}, 1424 (1998) and references therein. 
\bibitem{knob92} C. Knobler and R. Desai, Ann. Rev. Phys. Chem. {\bf
    43}, 207 (1992). 
\bibitem{sesh91} R. Seshadri and R. M. Westervelt,
  Phys. Rev. Lett. {\bf 66}, 2774 (1991).
\bibitem{murr92} C. M. Murray, in {\em Bond Orientational Order in
    Condensed Matter Systems}, edited by K. Strandburg (Springer,
  Berlin, 1992). 
\bibitem{zahn99} K. Zahn, R. Lenke and G. Maret,
  Phys. Rev. Lett. {\bf 82}, 2721 (1999). 
\bibitem{jast99} A. Jaster, Phys. Rev. E {\bf 59}, 2594 (1999). 
\bibitem{bagc96} K. Bagchi, H. C. Andersen and W. Swope, Phys. Rev. E
  {\bf 53}, 3794 (1996).  
\bibitem{some98} F. L. Somer, Jr., G. S. Canright and T. Kaplan,
  Phys. Rev. E  {\bf 58}, 5748 (1998).
\bibitem{park92} J. Park, T. C. Lubensky and F. C. MacKintosh,
  Europhys. Lett. {\bf 20}, 279  (1992) and references therein. 
\bibitem{evan96} R. M. L. Evans, Phys. Rev. E {\bf 53}, 935 (1996).
\bibitem{davi97} E. J. Davis, Aerosol Sci. Techn. {\bf 26}, 212
  (1997).
\bibitem{leid95} P. Leiderer, Z. Phys. B {\bf 98}, 303 (1995).
\bibitem{chia95} H. T. Chiang, V. S. Chen-White, R. Pindak and
  M. Seul, J. Phys. II {\bf 5}, 835 (1995). 
\bibitem{huan92} C. C. Huang, in {\em Bond Orientational Order in
    Condensed Matter Systems}, edited by K. Strandburg (Springer,
  Berlin, 1992). 
\bibitem{cele97} F. Celestini, F. Ercolessi and E. Tosatti,
  Phys. Rev. Lett. {\bf 78}, 3153 (1997). 
\bibitem{albr92} U. Albrecht and P. Leiderer, J. Low. Temp. Phys {\bf
    86}, 131 (1992).
\bibitem{leid97} P. Leiderer, in {\em Two-Dimensional Electron
    Systems}, edited by E. Y. Andrei (Kluwer Academic, Amsterdam, 1997).
\bibitem{salo81} M. M. Salomaa and G. A. Williams,
  Phys. Rev. Lett. {\bf 47}, 1730 (1981).
\bibitem{land}  L. D. Landau, E. M. Lifshitz and L. P. Pitaevskii,
  {\em  Electrodynamics of Continuous Media}, 2nd edition (Pergamon, New
  York, 1984). 
\bibitem{nels87} D. R. Nelson and L. Peliti, J. Phys. France {\bf 48},
  1085 (1987).
\bibitem{lenz01} P. Lenz and D. R. Nelson, Phys. Rev. Lett. {\bf 87},
  125703 (2001).
\bibitem{hild} U. Dierkes, S. Hildebrandt, A. K\"uster and O. Wohlrab,
{\em   Minimal Surfaces I}, (Springer, Berlin, 1992). 
\bibitem{nels83b} D. R. Nelson, in {\em Phase Transitions and Critical
    Phenomena}, Vol. 7, edited by C. Domb and J. Lebowitz (Academic, New
  York, 1983).
\bibitem{fran} T.~Frankel, {\em The Geometry of Physics} (Cambridge
  Univ. Press, Cambridge, 1997).
\bibitem{nels83} See, e.g., D. R. Nelson, Phys. Rev. B {\bf 28}, 5515
  (1983); see also S. Sachdev and D. R. Nelson, J. Phys. C {\bf 17},
  5473 (1984).
\bibitem{lube87} For the generalization to $p$-fold symmetric order
  parameters on the sphere see T. Lubensky and J. Prost, J. Phys. 
  France   {\bf 48}, 1085 (1987).
\bibitem{bowi00} M. J. Bowick, D. R. Nelson and A. Travesset,
  Phys. Rev. B {\bf 62}, 8738 (2000).
\bibitem{land2} L. D. Landau and E. M. Lifshitz, {\em Hydrodynamics}
  (Pergamon, New York, 1959). 
\bibitem{footnew} Corrections to this spectrum arising from the
non-linear term of Eq. (\ref{eq:2.14}) will be of higher order in
$r_{lm}$.  As can be seen from  Eqs. (\ref{eq:3.17}), (\ref{eq:3.18})
and (\ref{eq:3.25}) $\rho_l (\vec{v} \cdot  \nabla ) \vec{v}$ 
is of order $R_0 \omega^2r_{lm}^2$ whereas the leading terms 
$\rho_l \partial \vec{v}/\partial t$ and  $ \nabla  p$ are of order
$R_0 \omega^2r_{lm}$.
\bibitem{nels89} D. R. Nelson and F. Spaepen, Solid State Phys. {\bf
42}, 1 (1989).
\bibitem{chan}  S.~Chandrasekhar, {\em Hydrodynamic and Hydromagnetic
    Stability} (Dover,  New York, 1981).
\bibitem{fabe} T.~E.~Faber, {\em Fluid Dynamics for Physicists}
  (Cambridge Press, Cambridge, 1995).
\bibitem{grad} I. S. Gradshteyn and I. M. Rhyzik, {\it Table of Integrals,
  Series and Products}, 5th edition, (Academic Press, San Diego, 1994).
\bibitem{foot3} Here, we have assumed as appropriate for helium, that
the dielectric constants of the liquid and of its vapor phase are
comparable 
$\varepsilon_l \simeq \varepsilon_v \simeq \varepsilon \simeq 1$. 
The  general situation of
a charged sphere with dielectric constant $\varepsilon_1$ in
surrounding medium with dielectric constant $\varepsilon_2$ (with
$\varepsilon_1 \neq \varepsilon_2$) is more complicated. However,
in the limit of large $N$ (when the radius of the sphere becomes large
compared to interparticle spacing) one can in first order replace
$\varepsilon$ by $(\varepsilon_1+\varepsilon_2)/2$.
\bibitem{shik78} V. B. Shikin,   Pis'ma Zh. Eksp. Teor. Fiz. {\bf 27},
  44 (1978) [JETP Lett. {\bf 27}, 39 (1978)].
\bibitem{foot:4}
For simplicity, the $l=0$ mode has been excluded here. However, 
bubbles are stabilized by their finite
compressibility against purely radial oscillations. 
\bibitem{lenz03} P. Lenz, to be published.
\bibitem{schn84} M. B. Schneider, J. T. Jenkins and W. W. Webb, 
J. Phys. France {\bf 45}, 1457 (1984) and references therein. 
\bibitem{miln87} S. T. Milner and S. A. Safran,
Phys. Rev. A {\bf 36}, 4371 (1987).
\bibitem{foot:2.1}
The Reynolds number is 
${\mathrm R}{\mathrm e}=\rho v R_0/\eta$. A characteristic 
velocity is $v \simeq R_0/\tau$ with $\tau \simeq \eta R_0^3/\kappa$. Thus, 
${\mathrm R}{\mathrm e} \simeq \kappa \rho/\eta^2 R_0$. 
With $\kappa$ of the
order of 10 $k_BT$, $\eta/\rho \simeq 10^{-2}$cm$^2/$sec and 
$\rho \simeq 10^3$kg$/$m$^3$ one 
has ${\mathrm R}{\mathrm e} \simeq 10^{-5}$ for $R_0 \simeq 1
\mu$m. See also Ref. \cite{miln87}.
\bibitem{seif98} U. Seifert, Eur. Phys. J. B {\bf 8}, 405 (1999). 
\bibitem{happ} J. Happel and H. Brenner, {\em Low Reynolds Number
    Hydrodynamics} (Noordhoof Internat. Pub., Leiden, 1973).
\bibitem{pete85} M. A. Peterson, J. Math. Phys. {\bf 26}, 711 (1985).
\bibitem{mors} D. C. Morse and S. T. Milner, Phys. Rev. E 
{\bf 52}, 5918 (1995). 
\bibitem{foot:5.2} In fact Milner and Safran find that $r_{lm}$
diverges for $l=2$ in this case. Here, we do not take  this
instability into consideration. However, for a discussion of this
point see also \cite{seif97}. 
\bibitem{seif97} U. Seifert, Adv. Phys. {\bf 46}, 13 (1997).
\bibitem{davi87} F. David, E. Guitter and L. Peliti, J. Phys. France
{\bf 48}, 2059 (1987). 
\bibitem{barz94} R. Bar-Ziv and E. Moses, Phys. Rev. Lett. {\bf 73},
1392 (1994). 
\bibitem{nels95} P. Nelson, T. Powers and U. Seifert,
Phys. Rev. Lett. {\bf 74}, 3384 (1995).
\bibitem{zhan93} Z. Zhang, H. T. Davis and D. M. Kroll,
  Phys. Rev. E {\bf 48}, R651 (1993).
\bibitem{komu92} S.~Komura and R.~Lipowsky, 
J. Physique II {\bf 2}, 1563 (1992).
\bibitem{erdi00} S. Erdin and V. L. Pokrovsky, cond-mat/0008266 (2000).
\bibitem{davi} F. David, in {\em Statistical Mechanics of Membranes
    and Interfaces}, edited by D. Nelson, T. Piran and S. Weinberg
  (World Scientific, Singapore, 1988).
\bibitem{krey} E. Kreyszig, {\em Differential Geometry} (Dover, New
  York, 1991).
\bibitem{naka} M. Nakahara, {\em Geometry, Topology and Physics} (IOP
  Publishing, London, 1990).
\end{thebibliography}
\end{document}